\documentclass[%
reprint,
amsmath,
amssymb,
aps,
pra,
nofootinbib,
superscriptaddress,
floatfix]{revtex4-1}

\usepackage[pdftex]{graphicx}
\usepackage{hyperref}

\usepackage{bm}
\usepackage{braket}
\makeatletter

\let\MYcaption\@makecaption
\makeatother

\usepackage{subcaption}
\makeatletter
\let\@makecaption\MYcaption
\makeatother

\usepackage{color}
\usepackage[normalem]{ulem} 

\begin{document}

\title{Strongly parity-mixed superconductivity in Rashba-Hubbard model}

\author{Kosuke Nogaki}
\email[]{nogaki.kosuke.83v@st.kyoto-u.ac.jp}
\affiliation{%
  Department of Physics, Kyoto University, Kyoto 606-8502, Japan
}%

\author{Youichi Yanase}
\affiliation{%
  Department of Physics, Kyoto University, Kyoto 606-8502, Japan
}%
\affiliation{%
  Institute for Molecular Science, Okazaki 444-8585, Japan
}%

\date{\today}

\begin{abstract}
Heterostructures containing strongly correlated electron systems provide a platform to clarify interplay of electron correlation and Rashba spin-orbit coupling in unconventional superconductors. Motivated by recent fabrication of artificially-engineered heavy fermion superlattices and high-temperature cuprate superconductors, we conduct a thorough study on superconductivity in Rashba-Hubbard model. In contrast to previous weak coupling approaches, we employ fluctuation-exchange approximation to describe quantum critical magnetic fluctuations and resulting superconductivity. As a result, robust Fermi surfaces against magnetic fluctuations, incommensurate spin fluctuations, and a strongly parity-mixed superconducting phase are demonstrated in a wide range of electron filling from type-II van Hove singularity to half-filling. We also clarify impacts of type-II van Hove singularity on magnetic fluctuations and superconductivity. Whereas the $d_{x^2-y^2}$-wave pairing always dominant, subdominant spin-triplet pairing with either $p$-wave or $f$-wave symmetry shows a comparable magnitude, especially near the type-II van Hove singularity.
Our results resolve unsettled issues on strongly correlated Rashba systems and uncover candidate systems of nonreciprocal transport and topological superconductivity. 
\end{abstract}

\maketitle

\section{Introduction}
Recent development in engineering of two-dimensional crystalline electron systems has provided a new field in the superconducting research~\cite{Reyren1196,Ueno2008,Ye1193}. In particular, various phenomena unique to noncentrosymmetric superconductors have been observed in SrTiO$_3$ heterostrucures~\cite{Itahashieaay9120} and transition metal dichalcogenides~\cite{Saito2016,Lu1353,Xi2016}. Interplay of antisymmetric spin-orbit coupling (ASOC) and magnetic field has been focused on in these systems. On the other hand, fabrication of artificial superlattices containing strongly correlated electron systems, such as $\rm{CeCoIn_{5}/YbCoIn_{5}}$~\cite{mizukami2011extremely}, $\rm{CeCoIn_{5}/CeRhIn_{5}}$~\cite{naritsuka2018}, and $\rm{YbRhIn_{5}/CeCoIn_{5}/YbCoIn_{5}}$~\cite{naritsuka2017}, makes interplay of ASOC and strong electron correlations to be a fascionating topics. For instance, proposals of topological superconductivity~\cite{daido2016,daido2017,takasan2017,lu2018,yoshida2015,yoshida2017} and enhanced Edelstein effect~\cite{peters2018} shed light on potential impact of this topic on topological science and spintronics research. 

Bulk $\rm{CeCoIn_{5}}$ is one of the typical unconventional superconductors in the vicinity of the antiferromagnetic (AFM) quantum critical point~\cite{Petrovic_2001}. Non-Fermi liquid behaviors~\cite{Sidorov2002,Tayama2002,Nakajima2004,Kohori2001,Zaum2011} and $d_{x^2-y^2}$-wave superconductivity~\cite{an2010}, which are characteristic of magnetic criticality, have been established. 
Artificially-engineered superlattice containing a few layer $\rm{CeCoIn_{5}}$ naturally realizes two-dimensional $d_{x^2-y^2}$-wave superconductivity~\cite{Shimozawa_2016}. 
At the interface of heterostructures Rashba-type ASOC arises from polar inversion symmetry breaking~\cite{rashba1960properties}, and therefore, the superlattice containing heavy ions is expected to be affected by the Rashba ASOC. Depending on the superlattice structure, staggered or uniform Rashba ASOC appears, and accordingly, locally~\cite{Goh2012,Shimozawa2014} or globally~\cite{naritsuka2017} noncentrosymmetric superconductivity have been supported by experimental results for the heavy fermion superlattices~\cite{Shimozawa_2016}. Unique superconducting phases are expected to be realized there owing to the interplay of two-dimensional magnetic fluctuations and Rashba ASOC. 

Noncentrosymmetric structures can also be found in bulk materials. Indeed, vast studies of noncentrosymmetric superconductivity were triggered by the discovery of superconductivity in CePt$_3$Si~\cite{bauer_lec}. Furthermore, a recent experiment uncovered Rashba-type ASOC in a high-temperature cuprate superconductor $\rm{Bi_2Sr_2CaCu_2O_{8+\delta}}$~\cite{Gotlieb1271} whose crystal structure can be regarded as a naturally-formed superlattice. Controllability of artificial superlattices as well as spin-momentum locking uncovered in bulk materials generate renewed interest on noncentrosymmetric superconductivity in strongly correlated electron systems. 

Motivated by these considerations, we study superconductivity in the two-dimensional Rashba-Hubbard model. 
Although this model has been analyzed as a minimal model for strongly-correlated electron systems lacking inversion symmetry~\cite{yanase2007,yanase2008,Yokoyama2007,tada2008,takimoto2008,Shigeta2013,maruyama2015,greco2018,greco2019ferromagnetic,lu2018,wolf2020}, most of theoretical studies are based on weak-coupling approaches such as the perturbation theory or the random phase approximation (RPA)~\cite{yanase2007,yanase2008,Yokoyama2007,tada2008,takimoto2008,Shigeta2013,maruyama2015,greco2018,greco2019ferromagnetic}. 
In particular, analysis based on a theoretical method appropriate in quantum critical region has not been conducted.  
To clarify the superconducting phase stabilized by the interplay of critical magnetic fluctuations and Rashba ASOC, in this paper we adopted fluctuation exchange (FLEX) approximation which appropriately reproduces critical behaviors of self-consistent renormalization theory~\cite{moriya_review}. 

An electronic structure characteristic of the Rashba-Hubbard model is spin-splitting due to the Rashba ASOC and resulting type-II van Hove singularity which is positioned away from the time-reversal invariant momenta. We may expect unusual properties due to a large density of states when the Fermi surface is close to the van Hove singularity. Indeed, a recent theoretical study proposed the ferromagnetic (FM) spin fluctuation and spin-triplet $f$-wave superconductivity~\cite{greco2018,greco2019ferromagnetic}.
 In order to examine this proposal and to provide a thorough study of unconventional superconductivity in the Rashba-Hubbard model, we calculate the Fermi surfaces (FSs), magnetic susceptibility, and superconducting gap functions in a wide range of the filling. We show that FSs are robust against critical magnetic fluctuations in contrast to a previous theory~\cite{fujimoto2015deformation}. Furthermore, we show that strongly parity-mixed superconductivity with dominant $d_{x^2-y^2}$-wave pairing is robust in a whole parameter range in contrast to the proposal in Ref.~\cite{greco2018}. Interestingly, the parity mixing is enhanced near the van Hove singularity and the subdominant spin-triplet pairing has a magnitude comparable the spin-singlet pairing. We find signatures of the type-II van Hove singularity, such as the Lifshitz transition of FSs, strong instability to commensurate antiferromagnetic (CAFM) order, and the spin-triplet gap function changing from $p$-wave to $f$-wave. Then, the $f$-wave pairing is attributed not to the FM fluctuation but to the AFM fluctuation. Our study not only critically examines the previous works but also clarifies a mechanism of strongly parity-mixed superconducting states, which may be a platform of topological superconductivity~\cite{daido2016,daido2017,takasan2017,lu2018,yoshida2015,yoshida2017}, nonreciprocal electric transport~\cite{Itahashieaay9120,wakatsuki2018}, and fractional flux quanta~\cite{iniotakis2008}. 

The rest of the paper is constructed as follows. In Sec.~II, we introduce the Rashba-Hubbard model, and formulate the FLEX approximation and \'{E}liashberg equation for this model. In Sec.~III, we show the FSs and compare the noninteracting and interacting systems. The magnetic fluctuations and superconductivity are investigated in Secs.~IV and V, respectively. Correlation between the structures of magnetic susceptibilities and superconducting gap functions is revealed. A brief summary and discussions are provided in the last section, Sec.~VI.

\section{Model and method}

\subsection{Rashba-Hubbard model}
First, we introduce a Rashba-Hubbard model which describes
strongly correlated electron systems without inversion symmetry:
\begin{align}
  \label{num:full_hamiltonian}
  &\mathcal{H} = \mathcal{H}_0 + \mathcal{H}_{\rm{int}},
\\
  \label{num:free_hamiltonian}
  &\mathcal{H}_0 =
  \sum_{\bm{k} , \sigma} \varepsilon(\bm{k})
  c^{\dagger}_{\bm{k} \sigma} c_{\bm{k} \sigma}
  + \alpha \sum_{\bm{k}} \bm{g}(\bm{k}) \cdot \bm{S}(\bm{k}),
\\
  \label{num:int_hamiltonian}
  &\mathcal{H}_{\rm{int}} =
  U\sum_{i} n_{i \uparrow} n_{i \downarrow},
\end{align}
where
\begin{equation}
  \label{num:spin_op}
  \bm{S}(\bm{k})=\sum_{\bm{k} , \sigma, \sigma'}
  \bm{\sigma}_{\sigma\sigma'}
  c^{\dagger}_{\bm{k} \sigma} c_{\bm{k} \sigma'},
\end{equation}
is a momentum-selective spin operator,
$U$ represents on-site Coulomb repulsion,
$\bm{\sigma}$ are the Pauli matrices, and $c_{\bm{k} \sigma}$ ($c^{\dagger}_{\bm{k} \sigma}$) is an annihilation (creation) operator for an electron with momentum $\bm{k}$ and spin $\sigma$.
We consider the square lattice and assume a tight-binding energy dispersion,
\begin{equation}
  \label{num:band_dispersion}
  \varepsilon(\bm{k}) =
  -2t(\cos k_x+\cos k_y)
  +4t'\cos k_x \cos k_y -\mu,
\end{equation}
where $t$ and $t'$ represent first- and second-neighbor hopping integrals,
respectively. The chemical potential $\mu$ is included in $\varepsilon(\bm{k})$.
The second term in the free part of Hamiltonian, Eq.~\eqref{num:free_hamiltonian}, describes
the ASOC which appears in crystals lacking inversion symmetry. 
The g-vector, $\bm{g}(\bm{k})$, characterizes the structure of ASOC \cite{bauer_lec}, and it is Rashba type in polar noncentrosymmetric systems. We here assume a Rashba type g-vector represented as~\cite{yanase2007,yanase2008},
\begin{equation}
  \label{num:g_vector}
  \bm{g}(\bm{k}) = \left(-\frac{\partial \varepsilon(\bm{k})}{\partial k_y} ,
  \frac{\partial \varepsilon(\bm{k})}{\partial k_x} , 0 \right).
\end{equation}
The ASOC shows a form of the momentum-dependent Zeeman field. Therefore, spin degeneracy of the band is split by the ASOC, and bands with negative and positive helicity have distinct energy,
\begin{equation}
  \label{num:band_disp_chiral}
  E_\lambda(\bm{k}) = \varepsilon(\bm{k}) +\lambda
  \alpha |\bm{g}(\bm{k})|,
\end{equation}
where $\lambda = \pm$ is the helicity index.
Unless stated otherwise, we set a temperature $T=0.01$, $t'=0.3$, and $\alpha=0.5$ with a unit of energy $t=1$. 

\subsection{Green function and susceptibility}
The noninteracting Green functions for $U=0$ are expressed
by the $2\times2$ matrix form in the spin basis,
\begin{equation}
  \label{num:noint_green}
  \hat{G}^{(0)}(\bm{k},i\omega_n) = \left(i\omega_n\hat{I}-\varepsilon(\bm{k})\hat{I}
  -\alpha \bm{g}(\bm{k})\cdot\bm{\sigma}\right)^{-1},
\end{equation}
where $\omega_n=(2n+1)\pi T$ are fermionic Matsubara frequencies.
The noninteracting Green functions in the helicity basis,
\begin{equation}
  \label{num:nonint_chiral_green}
  G^{(0)}_\lambda(\bm{k},i\omega_n) = \frac{1}{i\omega_n-\varepsilon(\bm{k})
    -\lambda\alpha |\bm{g}(\bm{k})|},
\end{equation}
are obtained by unitary transformation with $\hat{V}$ which diagonalizes $\mathcal{H}_0$ [Eq.~(\ref{num:band_disp_chiral})],
\begin{align}
  \hat{V}^\dagger\mathcal{H}_0\hat{V} = \left(
  \begin{array}{cc}
      E_+(\bm{k}) & 0           \\
      0           & E_-(\bm{k}) \\
    \end{array}
  \right).
\end{align}
These Green functions are connected by the following relationship,
\begin{equation}
  \hat{G}^{(0)}(\bm{k},i\omega_n) = \sum_{\lambda=\pm}
  \frac{1}{2} \left( \hat{I}+\lambda\frac{\bm{g}}{|\bm{g}|}
  \cdot \bm{\sigma} \right) G^{(0)}_\lambda(\bm{k},i\omega_n).
\end{equation}

In the interacting case $U\neq0$, the dressed Green functions contain a self-energy, $\hat{\Sigma}(\bm{k},i\omega_n)$, 
\begin{align}
  \label{num:green_function_dressed}
  \hat{G}(\bm{k},i\omega_n) & = \left(i\omega_n\hat{I}-\varepsilon(\bm{k})\hat{I}
  -\alpha \bm{g}(\bm{k})\cdot\bm{\sigma}-\hat{\Sigma}(\bm{k},i\omega_n)\right)^{-1}.
\end{align}
Within the FLEX approximation, the self-energy is expressed with use of an effective interaction, $\hat{\Gamma}^n(\bm{k},i\nu_n)$, as
\begin{align}
  \label{num:self_energy}
  &\Sigma_{\sigma\sigma'}  (\bm{k},i\omega_n) 
                          = T \sum_{\bm{q},i\nu_n}
  \Gamma^n_{\sigma\xi\sigma'\eta}(\bm{q},i\nu_n)G_{\xi\eta}(\bm{k}-\bm{q},i\omega_n-i\nu_n),
\end{align}
and the effective interaction is given by
\begin{equation}
  \label{num:effective_interaction}
  \hat{\Gamma}^n(\bm{k},i\nu_n)
  = \hat{U}\hat{\chi}(\bm{k},i\nu_n)\hat{U},
\end{equation}
where
\begin{align}
  \hat{U} = \left(
  \begin{array}{cccc}
      0  & 0 & 0 & -U \\
      0  & U & 0 & 0  \\
      0  & 0 & U & 0  \\
      -U & 0 & 0 & 0  \\
    \end{array}
  \right),
\end{align}
$\hat{\chi}(\bm{k},i\nu_n)$ is the generalized susceptibility,
and $i\nu_n$ are bosonic Matsubara frequencies.
We introduce the bare susceptibility
\begin{align}
  \label{num:bare_suscep}
  &\chi^{(0)} _{\sigma_1\sigma_2\sigma_3\sigma_4}(\bm{q},i\nu_n) \notag         \\
             & = -T\sum_{\bm{k},i\omega_n}G_{\sigma_1\sigma_3}(\bm{k},i\omega_n)
  G_{\sigma_4\sigma_2}(\bm{k}-\bm{q},i\omega_n-i\nu_n),
\end{align}
and compute the generalized susceptibility by
\begin{align}
  \label{num:gener_suscep}
  \hat{\chi}(\bm{q},i\nu_n) =
  \left[
    \hat{I}-\hat{\chi}^{(0)}(\bm{q},i\nu_n)\hat{U}
    \right]^{-1}
  \hat{\chi}^{(0)}(\bm{q},i\nu_n).
\end{align}
According to Eqs.~(\ref{num:green_function_dressed})-(\ref{num:gener_suscep}),
$\hat{G},\hat{\Sigma},\hat{\Gamma}^n,\hat{\chi}^{(0)},\hat{\chi}$
depend on each other, and therefore, we self-consistently determine these functions.
As a consequence of the self-consistent condition, the FLEX approximation is a conserving approximation in which several conservation laws are satisfied in the framework of the Luttinger-Ward theory \cite{Luttinger1960,Luttinger1960_2,Baym1961,Baym1962}.

Introducing the vector representation of the self-energy
\begin{equation}
  \hat{\Sigma} = \Sigma_0\hat{I} + \bm{\Sigma}\cdot\bm{\sigma},
\end{equation}
and carrying out analytic continuation,
we represent the renormalized retarded Green functions as
\begin{equation}
  \hat{G}^{\rm{R}}(\bm{k},\omega) = \left(\omega\hat{I}-\varepsilon'(\bm{k},\omega)\hat{I}
  -\alpha \bm{g}'(\bm{k},\omega)\cdot\bm{\sigma}\right)^{-1},
\end{equation}
where $\varepsilon' = \varepsilon+\Sigma^{\rm{R}}_0$ and $\alpha \bm{g}'=\alpha \bm{g} + \rm{Re} \bm{\Sigma}^{\rm{R}}$.
Since $\rm{Im} \bm{\Sigma}^{\rm{R}}$ is proportional to $T^2$ in a Fermi liquid state, we dropped it
and obtain
\begin{equation}
  \hat{G}^{\rm{R}}(\bm{k},\omega) = \sum_{\lambda=\pm}
  \frac{1}{2} \left( \hat{I}+\lambda\frac{\bm{g}'}{|\bm{g}'|}
  \cdot \bm{\sigma} \right) G^{\rm{R}}_\lambda(\bm{k},\omega),
\end{equation}
where
\begin{equation}
  \label{num:renom_chiral_green}
  G^{\rm{R}}_\lambda(\bm{k},\omega) = \frac{1}{\omega-\varepsilon'(\bm{k},\omega)
    -\lambda\alpha |\bm{g}'(\bm{k},\omega)|}.
\end{equation}
From Eq.~(\ref{num:renom_chiral_green}), we determine FSs of interacting systems by solving
\begin{equation}
  \varepsilon'(\bm{k},0)
  -\lambda\alpha |\bm{g}'(\bm{k},0)|=0.
\end{equation}
In this calculation, a static function $A(\bm{q},0)$ is evaluated by an approximation justified at low temperatures,
\begin{equation}
  A(\bm{q},0) \simeq \frac{A(\bm{q},i\pi T)+A(\bm{q},-i\pi T)}{2}.
\end{equation}

\subsection{Linearized \'{E}liashberg equation}
To investigate superconductivity, we numerically solve
the linearized \'{E}liashberg equation which is given by
\begin{align}
  \lambda\Delta_{\sigma\sigma'}(k) & =
  T\sum_{k'} \Gamma_{\sigma s_1 s_2 \sigma'}(k-k')F_{s_1s_2}(k'),                        \\
  F_{s_1s_2}(k')                   & = G_{s_1s_3}(k')\Delta_{s_3s_4}(k')G_{s_2s_4}(-k'),
\end{align}
where $\hat{\Delta}$ is the gap function and $\hat{\Gamma}$ is
obtained by
\begin{align}
  \hat{\Gamma}(k-k') & = \hat{U} +\hat{\Gamma}^n(k-k').
\end{align}
Here we adopted abbreviated notation $k=(\bm{k},i\omega_n)$. 
Evaluating $\lambda$, eigenvalues of
the linearized \'{E}liashberg equation, we determine the
critical temperature $T_c$ from the criterion $\lambda=1$. 

Even when $\lambda \ne 1$, we can identify the leading superconducting instability 
by comparing $\lambda$ for irreducible representations of a given point group. 
Since the point group symmetry of the Rashba-Hubbard model is $C_{4v}$,
the gap function can be classified into irreducible representations of $C_{4v}$.
We numerically calculate eigenvalues for each irreducible representations and 
conclude that the $B_1$ representation gives the largest eigenvalue $\lambda$ in the whole parameter range. 

\section{Fermi surfaces}
\label{sec:spin_suscep}
\begin{figure}[tbp]
  \begin{minipage}[b]{0.24\linewidth}
    \centering
    \includegraphics[keepaspectratio, scale=0.08]{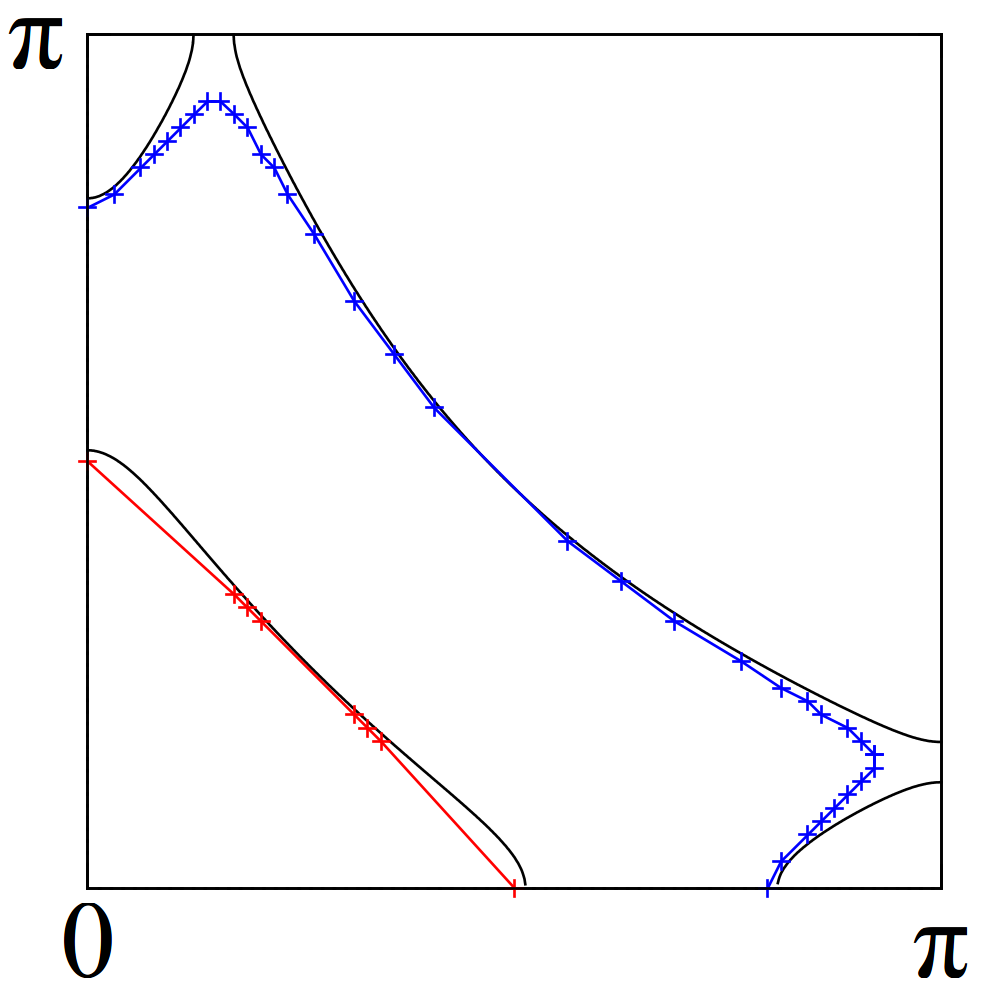}
    \subcaption{n=0.65}
    \label{fig:FS_n=0.65}
  \end{minipage}
  \begin{minipage}[b]{0.24\linewidth}
    \centering
    \includegraphics[keepaspectratio, scale=0.08]{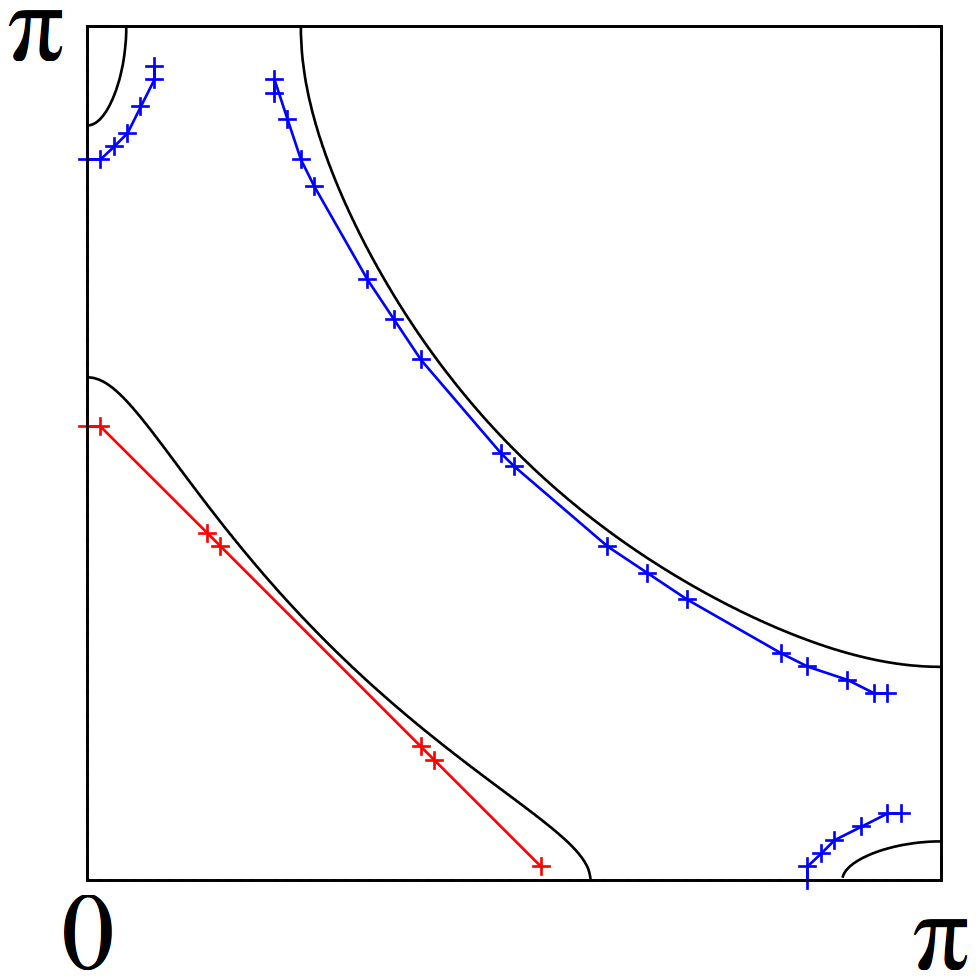}
    \subcaption{n=0.75}
    \label{fig:FS_n=0.75}
  \end{minipage}
  \begin{minipage}[b]{0.24\linewidth}
    \centering
    \includegraphics[keepaspectratio, scale=0.08]{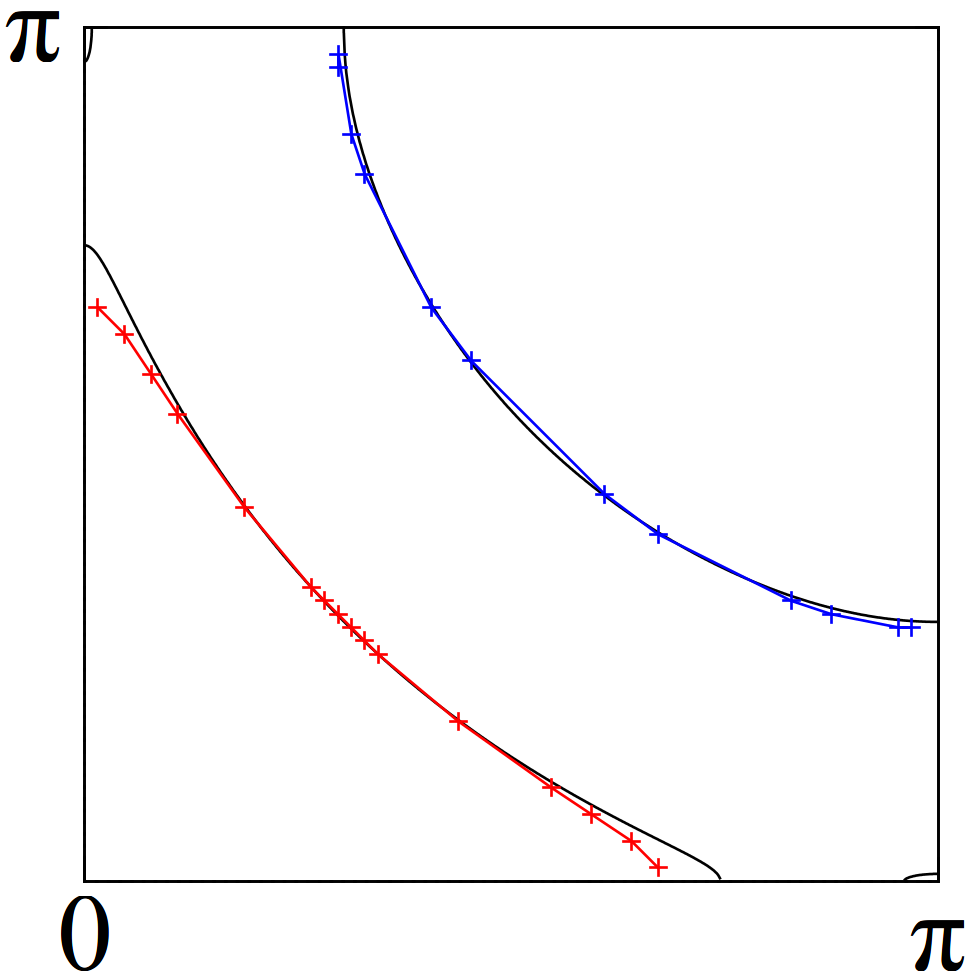}
    \subcaption{n=0.85}
    \label{fig:FS_n=0.85}
  \end{minipage}
  \begin{minipage}[b]{0.24\linewidth}
    \centering
    \includegraphics[keepaspectratio, scale=0.08]{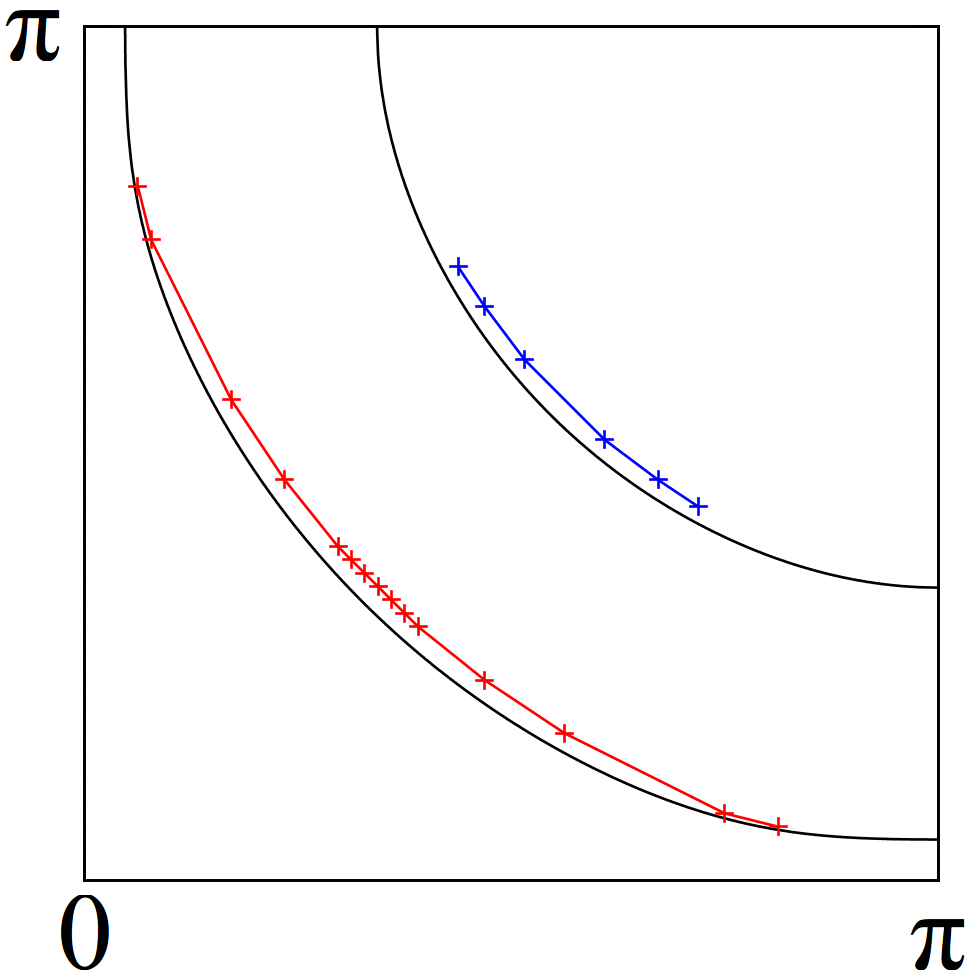}
    \subcaption{n=0.95}
    \label{fig:FS_n=0.95}
  \end{minipage} \\
  \caption{(Color online) Fermi surfaces of noninteracting systems ($U=0$) and interacting systems ($U\neq0$) for various carrier densities $n$. For interacting systems, red and blue lines show the FSs of positive and negative helicity bands, respectively. The FSs of noninteracting cases are plotted by black lines. 
  On-site Coulomb interaction is assumed to be $U=2.4$ for $n=0.65$ and $U=5$ for the others.}
  \label{fig:FS}
\end{figure}

As is known by vast previous works \cite{YANASE20031,Kuroki2006}, 
the topology and shape of FSs are closely related to magnetic fluctuations and superconductivity. 
Therefore, we begin with discussions about the FSs of interacting systems. In Fig.~\ref{fig:FS}(\subref{fig:FS_n=0.65})-(\subref{fig:FS_n=0.95}) we compare
the FSs in the noninteracting systems ($U=0$) with those in the interacting systems ($U \ne 0$). The electron filling is varied from $n=0.65$ near the type-II van Hove singularity to $n=0.95$ near half-filling. 
Red (blue) lines show interacting FSs of positive (negative) helicity bands, while the noninteracting FSs are plotted by black lines. 
Over a wide range of filling, the FSs of interacting systems almost coincide with those of noninteracting systems. 

For the effect of electron correlations on spin splitting in noncentrosymmetric systems, qualitatively different conclusions have been obtained in the previous studies. Ref.~\cite{maruyama2015} showed that the FSs are almost unchanged when the g-vector is represented in terms of the velocity as in Eq.~\eqref{num:g_vector}.
On the other hand, significant deformation of FSs by critical magnetic fluctuations is claimed in Ref.~\cite{fujimoto2015deformation}. 
Our numerical results for $n=0.75$, $0.85$, and $0.95$ support the former. Although Ref.~\cite{maruyama2015} conducted third-order perturbation theory for the Rashba-Hubbard model, we have shown that the FSs are robust even in the presence of critical magnetic fluctuations. Drastic change of FSs predicted in Ref.~\cite{fujimoto2015deformation} is not observed in our calculations. 
On the other hand, for $n=0.65$ the Fermi level is close to the type-II van Hove singularity, and then Lifshitz transition is caused by electron correlations. 

\section{Magnetic fluctuations}
\label{sec:mag}
\begin{figure}[tbp]
  \begin{minipage}[b]{0.24\linewidth}
    \begin{center}
    \includegraphics[keepaspectratio, scale=0.08]{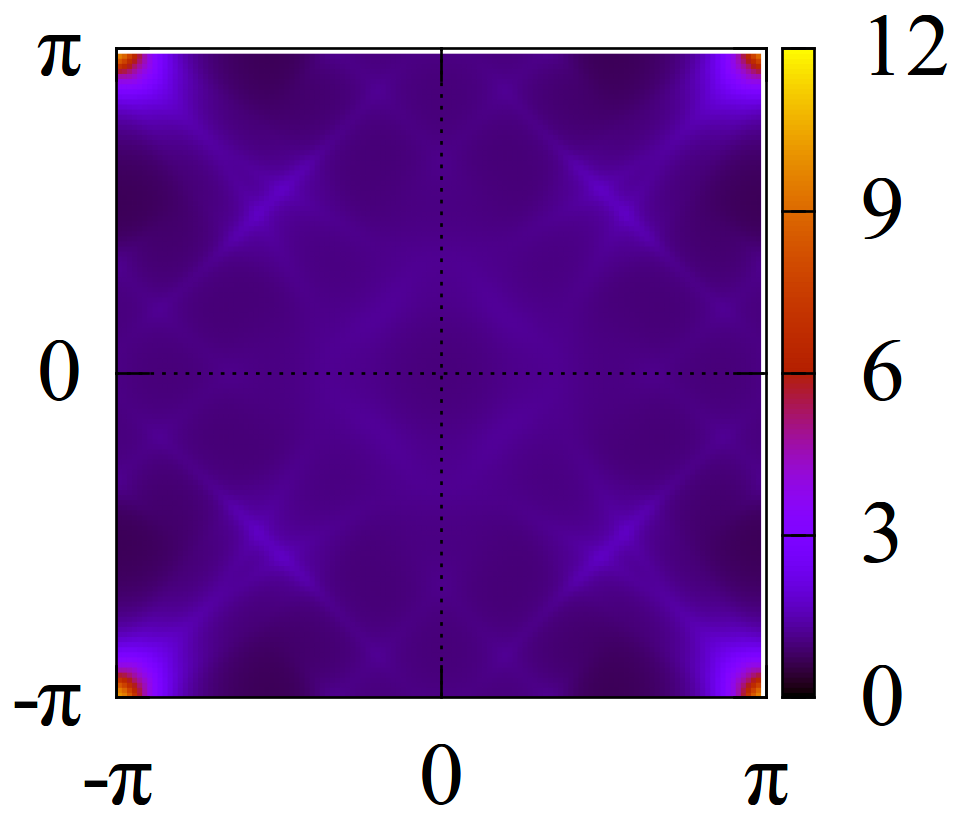}
    \subcaption{n=0.65}
    \label{fig:lki_n=0.65}
        \end{center}
  \end{minipage}
  \begin{minipage}[b]{0.24\linewidth}
    \begin{center}
    \includegraphics[keepaspectratio, scale=0.08]{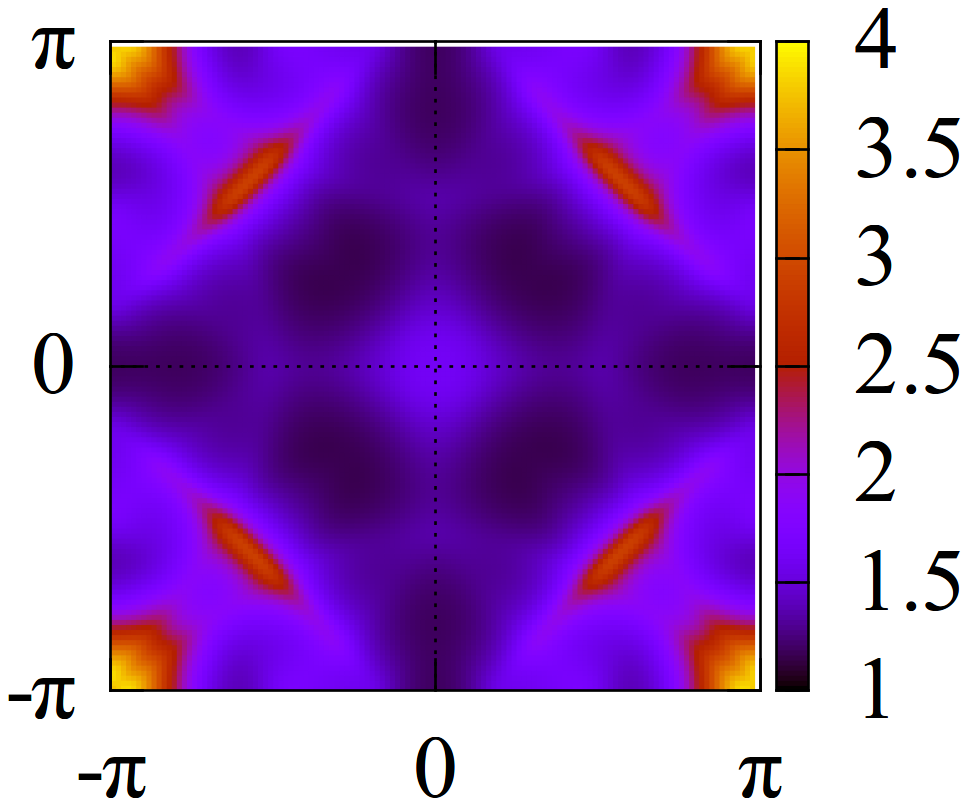}
    \subcaption{n=0.75}
    \label{fig:lki_n=0.75}
        \end{center}
  \end{minipage}
  \begin{minipage}[b]{0.24\linewidth}
    \begin{center}
    \includegraphics[keepaspectratio, scale=0.08]{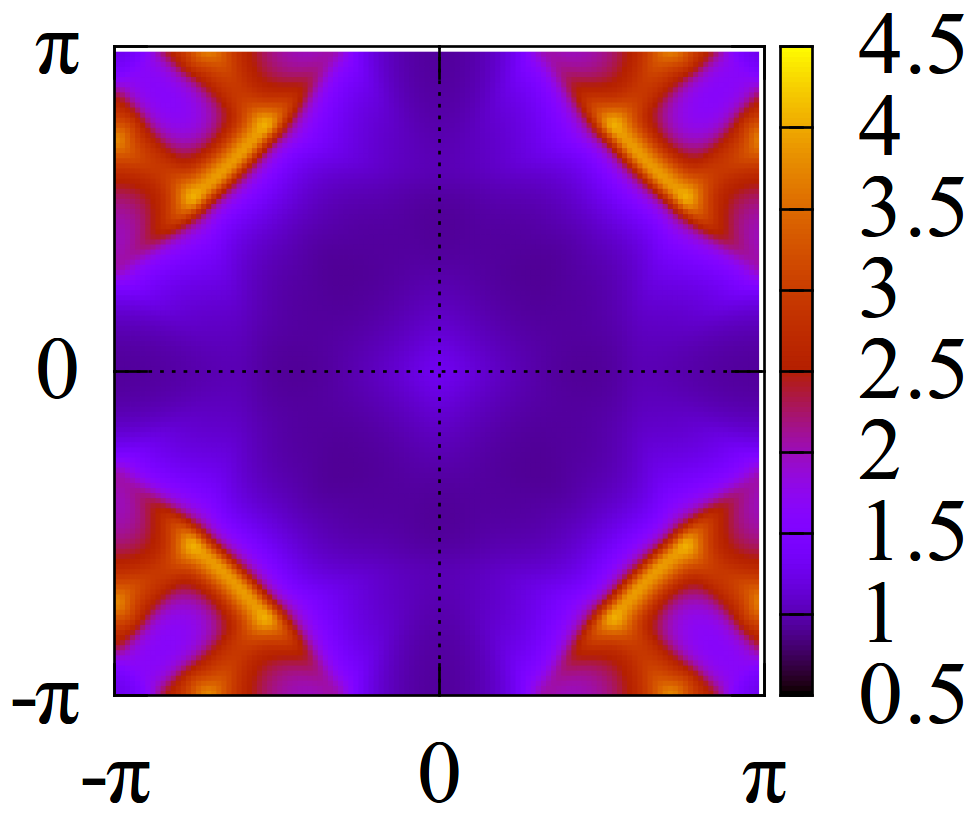}
    \subcaption{n=0.85}
    \label{fig:lki_n=0.85}
     \end{center}
  \end{minipage}
  \begin{minipage}[b]{0.24\linewidth}
    \begin{center}
    \includegraphics[keepaspectratio, scale=0.08]{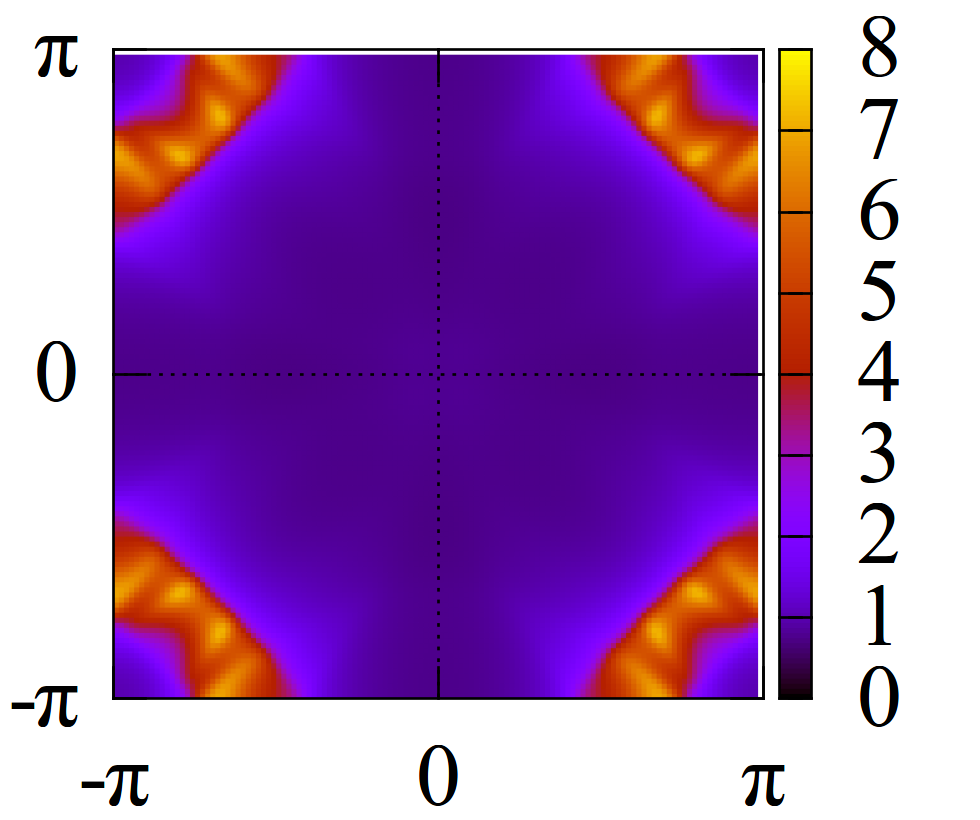}
    \subcaption{n=0.95}
    \label{fig:lki_n=0.95}
        \end{center}
  \end{minipage}
  \begin{minipage}[b]{0.24\linewidth}
    \begin{center}
    \includegraphics[keepaspectratio, scale=0.08]{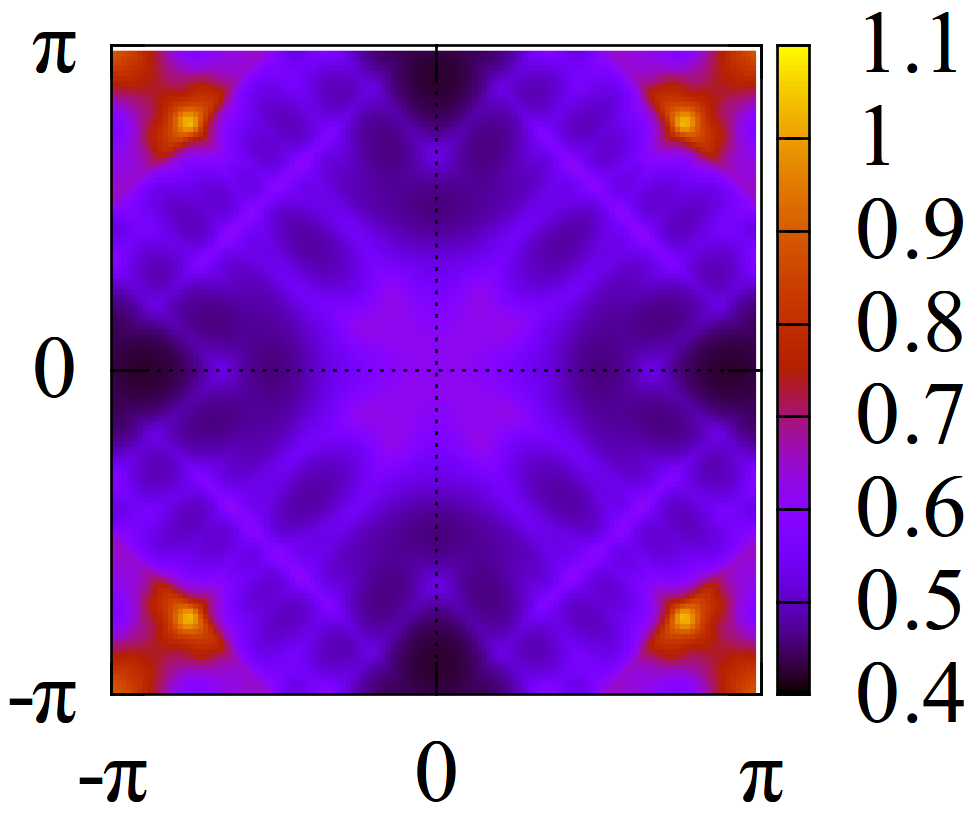}
    \subcaption{n=0.65}
    \label{fig:tki_n=0.65}
        \end{center}
  \end{minipage}
  \begin{minipage}[b]{0.24\linewidth}
    \begin{center}
    \includegraphics[keepaspectratio, scale=0.08]{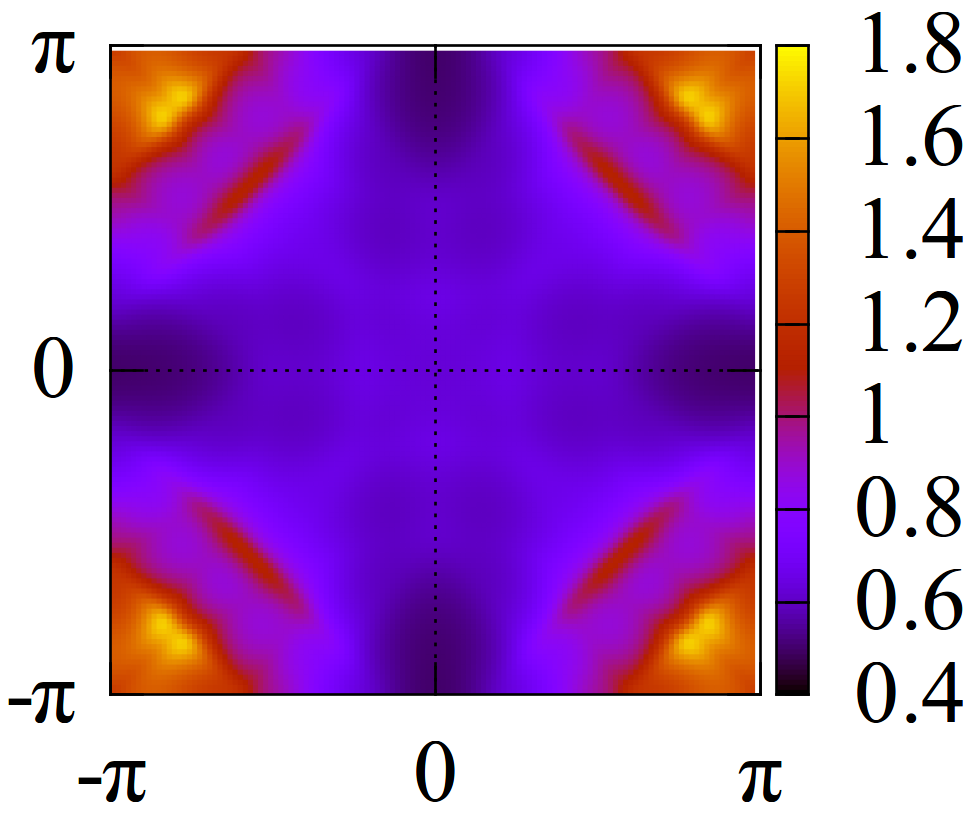}
    \subcaption{n=0.75}
    \label{fig:tki_n=0.75}
        \end{center}
  \end{minipage}
  \begin{minipage}[b]{0.24\linewidth}
    \begin{center}
    \includegraphics[keepaspectratio, scale=0.08]{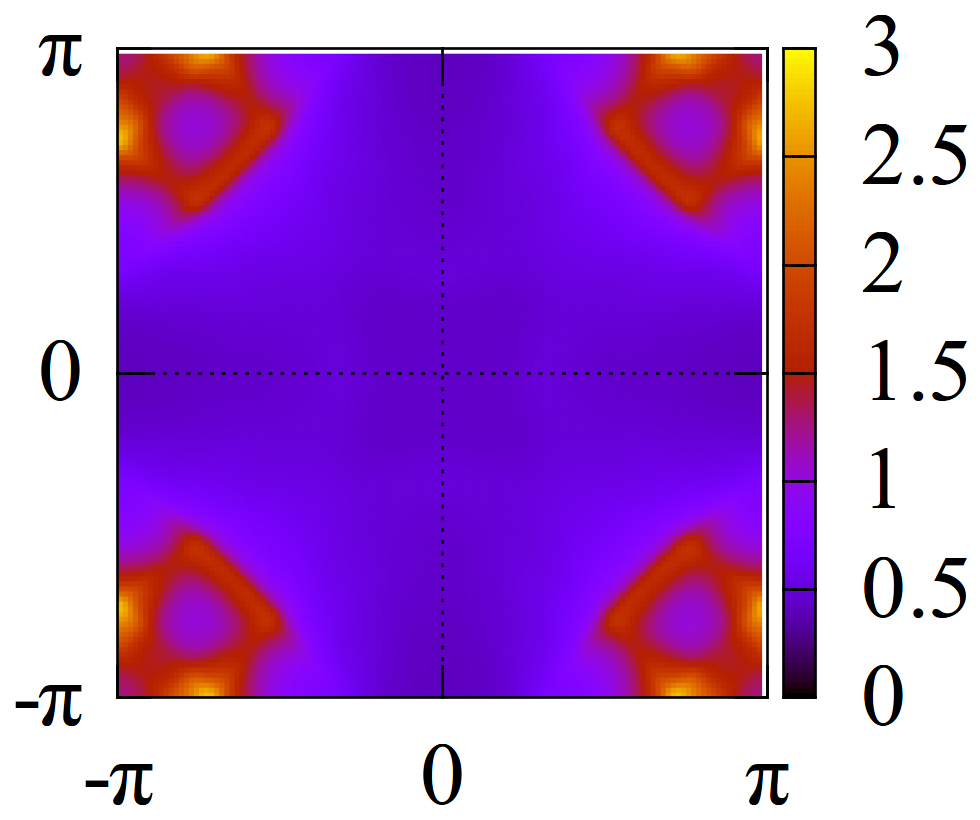}
    \subcaption{n=0.85}
    \label{fig:tki_n=0.85}
        \end{center}
  \end{minipage}
  \begin{minipage}[b]{0.24\linewidth}
    \begin{center}
    \includegraphics[keepaspectratio, scale=0.08]{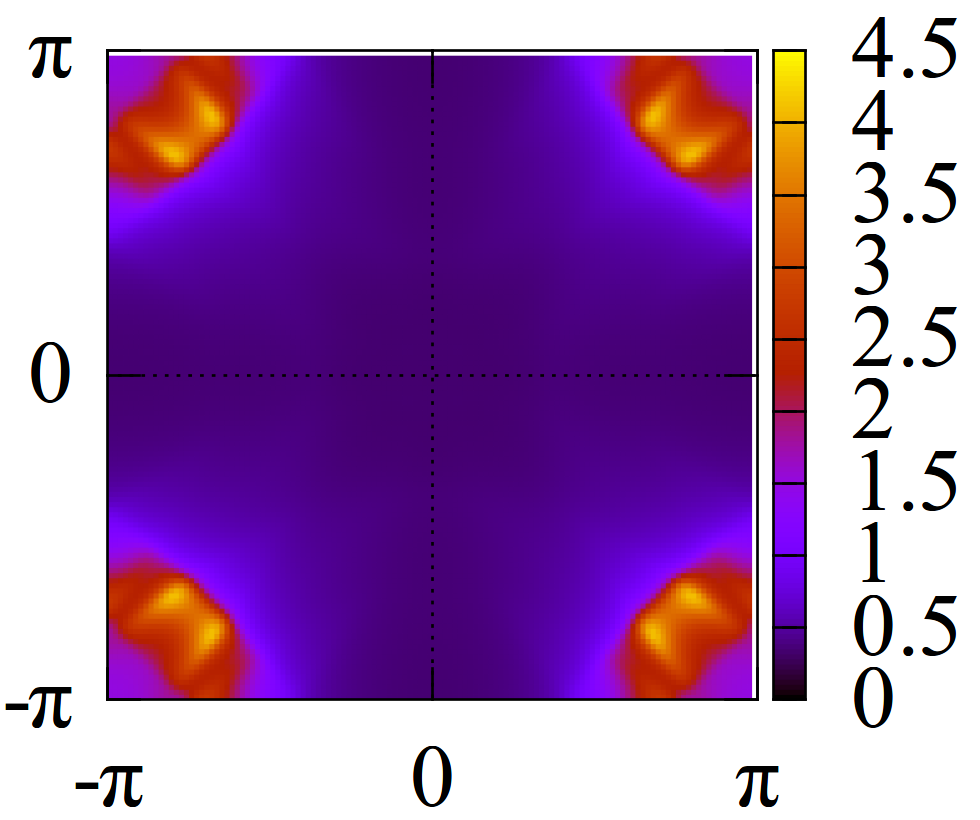}
    \subcaption{n=0.95}
    \label{fig:tki_n=0.95}
        \end{center}
  \end{minipage}
  \caption{(Color online) Momentum dependence of spin susceptibilities.
    (a)-(d) Longitudinal spin susceptibility and (e)-(h) transverse spin susceptibilities for the filling, $n=0.65$, $0.75$, $0.85$, and $0.95$.
    We choose $U=2.4$ for $n=0.65$ and $U=5$ for the others.}
  \label{fig:FS_sus}
\end{figure}

Next, we discuss magnetic fluctuations. 
Dynamical spin susceptibility tensor is given by the generalized susceptibility as
\begin{align}
  \chi^{\mu\nu}(\bm{q},i\nu_n)=\sum_{s_1s_2s_3s_4}\sigma^\mu_{s_1s_2}\chi_{s_2s_1s_3s_4}(\bm{q},i\nu_n)\sigma^\nu_{s_3s_4}.
\end{align}
We illustrate static longitudinal spin susceptibility, $\chi^{zz}$, 
and transverse spin susceptibility, $\chi^{-+}$, at $\nu_n=0$ in Fig.~\ref{fig:FS_sus} for various fillings.
Although $\chi^{zz}$ and $\chi^{-+}$ are equivalent in the absence of the spin-orbit coupling, magnetic anisotropy is induced by the Rashba ASOC in this model. 

The transverse spin susceptibility shows qualitatively similar momentum dependence for all fillings [Figs.~\ref{fig:FS_sus}(\subref{fig:tki_n=0.65})-(\subref{fig:tki_n=0.95})]. 
The incommensurate antiferromagnetic (IAFM) fluctuation appears in a wide range of filling, $0.65 \lesssim n \lesssim 0.95$.
A weak CAFM fluctuation with the modulation vector $\bm{Q}=(\pi,\pi)$ also develops at $n=0.65$ near the type-II van Hove singularity.

On the other hand, we observe a significant enhancement of the CAFM fluctuation in the longitudinal spin susceptibility at $n=0.65$ [Fig.~\ref{fig:FS_sus}(\subref{fig:lki_n=0.65})], whereas the IAFM fluctuation is dominant for $0.75 \lesssim n \lesssim 0.95$ [Figs.~\ref{fig:FS_sus}(\subref{fig:lki_n=0.75})-(\subref{fig:lki_n=0.95})]. 
The longitudinal spin susceptibility not only reveals the AFM fluctuations but also implies the FM spin fluctuation when the Fermi level is close to the type-I or type-II van Hove singularity~\cite{greco2019ferromagnetic}, as we see a weak peak at $\bm{Q}=(0,0)$ [Figs.~\ref{fig:FS_sus}(\subref{fig:lki_n=0.75}) and \ref{fig:FS_sus}(\subref{fig:lki_n=0.85}) as well as Fig.~\ref{fig:U=2.4}(\subref{fig:lki_n=0.85_U=2.4})]. However, consistent with the previous analysis based on the RPA~\cite{greco2019ferromagnetic}, the FM fluctuation is weakened in the strong coupling region. Indeed, we see only a weak signature of the FM spin fluctuation. As we show later, this FM spin fluctuation is almost unrelated to the superconductivity.  

For all fillings in Fig.~\ref{fig:FS_sus}, the longitudinal spin correlation is stronger than the transverse one. Thus, the Ising-type AFM spin fluctuation with dominant {\it c}-axis component is implied. The magnetic anisotropy is enhanced when the filling $n$ is decreased and the Fermi level approaches to the type-II van Hove singularity. 

Growth in the maximum value of the longitudinal spin susceptibility, $\chi^{zz}(\bm{Q},0)$, at $n=0.65$ suggests that the system is in the vicinity of the CAFM order. 
It should be noticed that in Fig.~\ref{fig:FS_sus} we choose $U=2.4$ for $n=0.65$ while $U=5$ for other fillings. This is because $U=5$ is larger than the critical interaction for the AFM order at $n=0.65$.
In fact, the critical interaction is approximately $U_{\rm c}=3.3$ for $n=0.65$, whereas $U_c > 6$ for other fillings. Thus, it is indicated that the AFM order develops when the FSs are close to the type-II van Hove singularity. Such filling dependence of magnetic fluctuations is qualitatively different from the conventional Hubbard model without the Rashba ASOC. In the ordinary Hubbard model, the magnetic correlations are enhanced near the half-filling. On the other hand, the magnetic correlations are mainly determined by the type-II van Hove singularity in the Rashba-Hubbard model with a large ASOC.

\section{Superconductivity}
\begin{figure}[tbp]
 \begin{minipage}[b]{\linewidth}
    \begin{center}
        \includegraphics[keepaspectratio, scale=0.18]{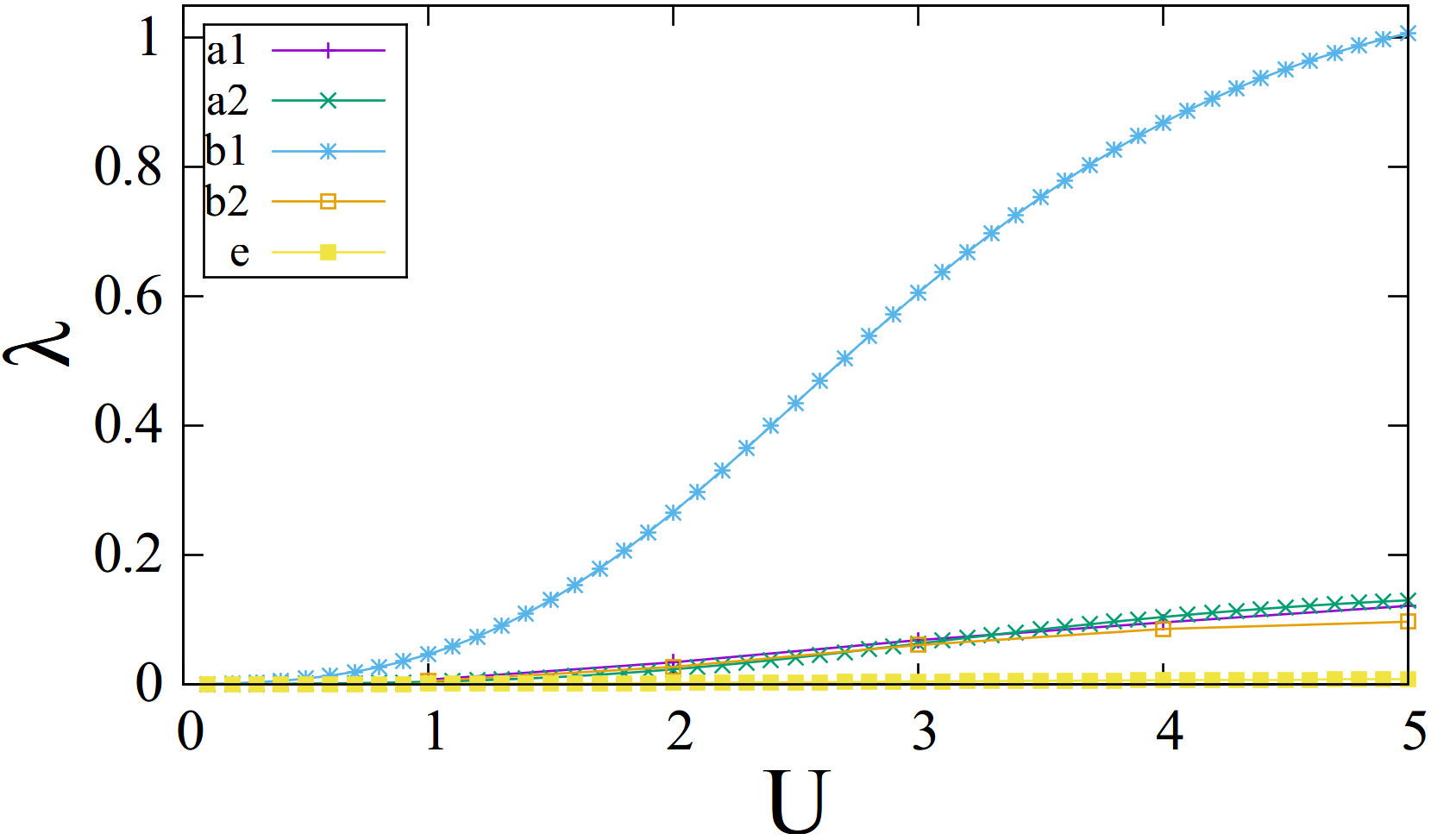}
        \subcaption{$U$-dependence}
        \label{fig:eigen_n=0.85}
    \end{center}
 \end{minipage} \\
 \hspace{10mm}
 \begin{minipage}[b]{\linewidth}
    \begin{center}
        \includegraphics[keepaspectratio, scale=0.18]{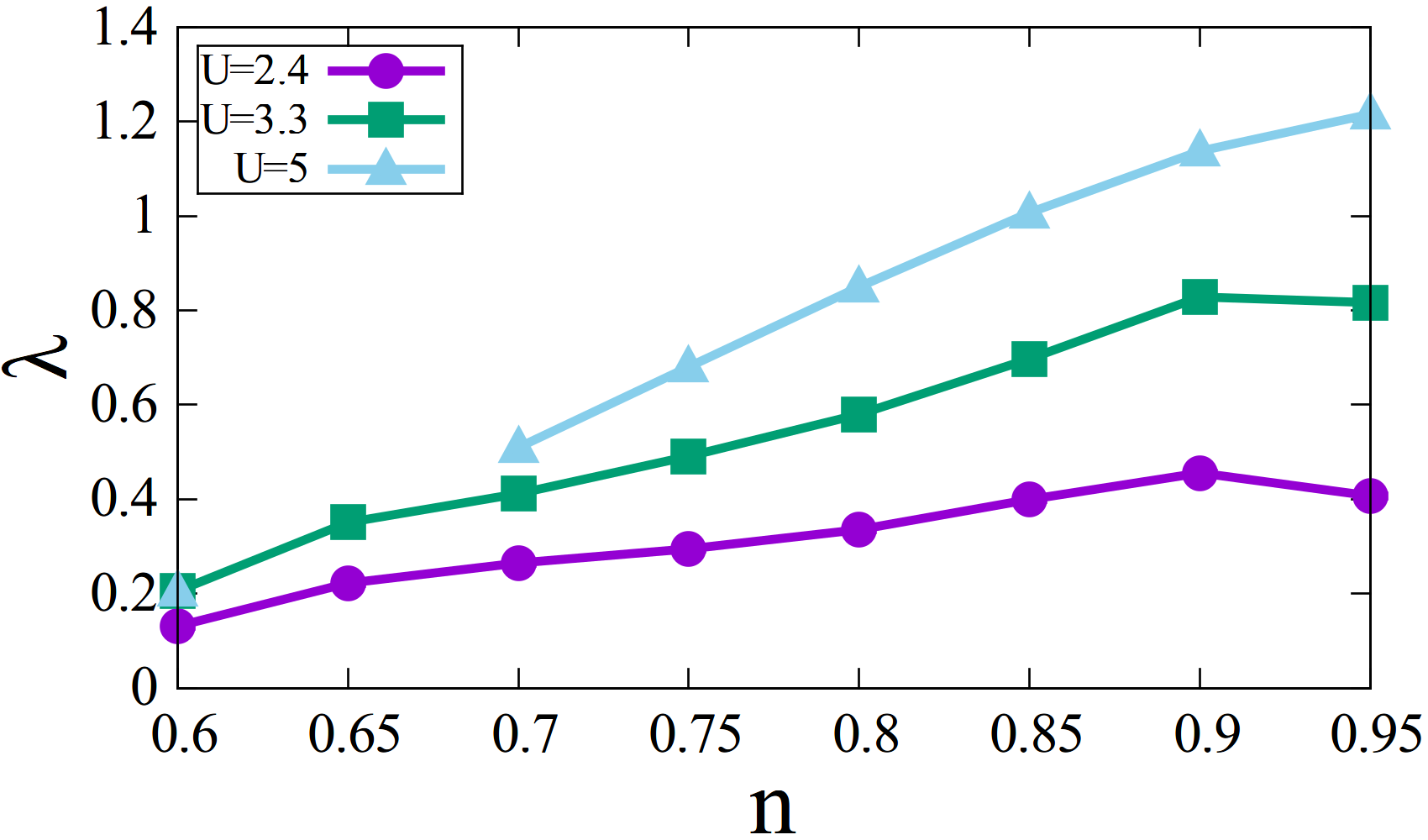}
        \subcaption{Filling dependence}
        \label{fig:lambda_filling}
    \end{center}
 \end{minipage}
  \caption{(Color online) (\subref{fig:eigen_n=0.85}) Eigenvalues of the linearized \'{E}liashberg equation $\lambda$ as a function of $U$ for the  $A_{1}$, $A_{2}$, $B_{1}$, $B_{2}$, and $E$ superconducting states. We assume $n=0.85$.
  (\subref{fig:lambda_filling}) Filling dependence of the eigenvalue for the most stable $B_{1}$ superconducting state. We choose $U=2.4$, $3.3$, and $5$.
  }
  \label{fig:eigen}
\end{figure}

Here we study the superconductivity. 
Superconducting phases are classified based on irreducible representations of the point group symmetry of the system, that is, $C_{4v}$. We calculate eigenvalues of the linearized \'{E}liashberg equation for all the irreducible representations, $A_{1}$, $A_{2}$, $B_{1}$, $B_{2}$, and $E$. 
For instance, Fig.~\ref{fig:eigen}(\subref{fig:eigen_n=0.85}) shows the interaction dependence of $\lambda$ and reveals that the superconducting phase of $B_1$ representation is the most stable. We confirmed that the $B_1$ superconducting phase is stable in the whole filling range investigated in this paper, that is, from half-filling to the type-II van Hove singularity. 

Fig.~\ref{fig:eigen}(\subref{fig:lambda_filling}) shows the filling dependence of the maximum eigenvalue $\lambda$ for the $B_1$ representation. The results suggest superconductivity with a high transition temperature near the half-filling, whereas the transition temperature decreases in the low-filling region. It should be noticed that the magnetic fluctuation grows near the type-II van Hove singularity more strongly than near the half-filling (Fig.~\ref{fig:FS_sus}). Our results indicate a weak tendency to superconductivity near the type-II van Hove singularity in spite of a strong instability to the CAFM order. This is partly because the magnetic fluctuation is significantly localized in the momentum space: $\chi^{zz}(\bm{q},0)$ shows a sharp peak around the commensurate wave vector $\bm{q} = \bm{Q}$. It makes total weight of the spin fluctuation, $\int {\rm d}\bm{q} \chi^{zz}(\bm{q},0)$, to be small. A strong magnetic anisotropy also favors magnetic order rather than superconductivity. Because both longitudinal and transverse spin fluctuations mediate an attractive interaction for spin-singlet pairing~\cite{YANASE20031}, an isotropic spin fluctuation may give rise to higher superconducting transition temperatures than the Ising spin fluctuation.

\begin{figure}[tbp]
  \begin{minipage}[b]{0.24\linewidth}
    \begin{center}
    \includegraphics[keepaspectratio, scale=0.08]{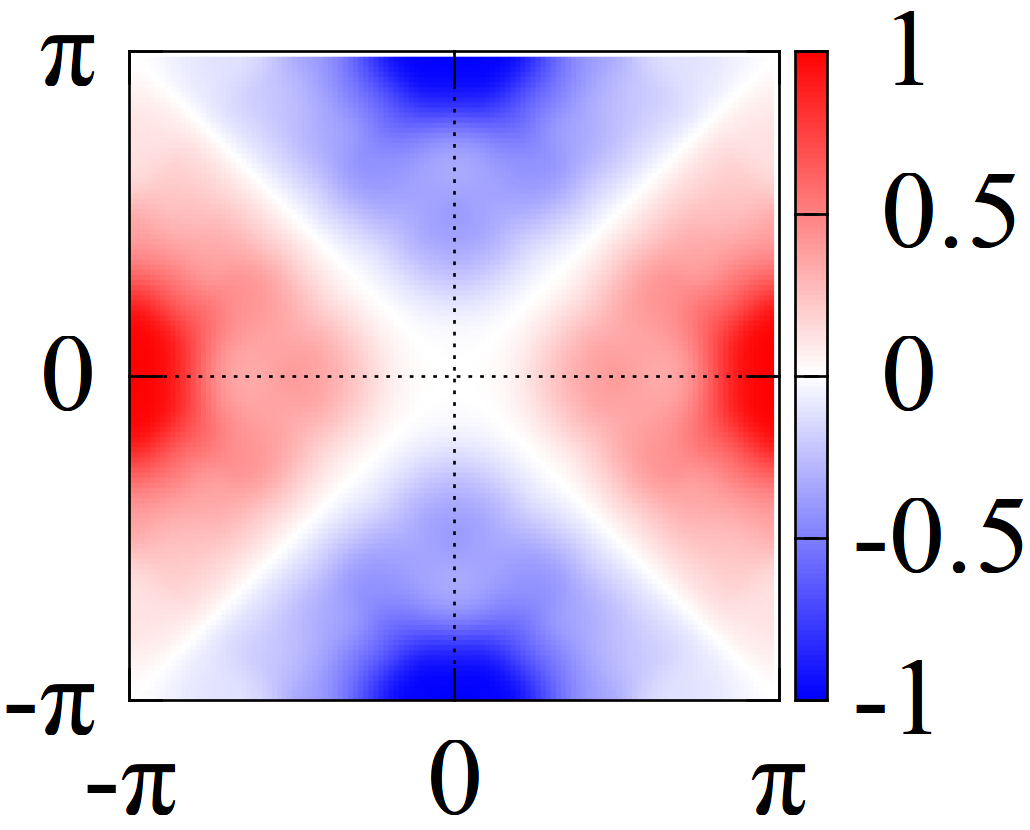}
    \subcaption{n=0.65}
    \label{fig:phi_n=0.65}
        \end{center}
  \end{minipage}
  \begin{minipage}[b]{0.24\linewidth}
    \begin{center}
    \includegraphics[keepaspectratio, scale=0.08]{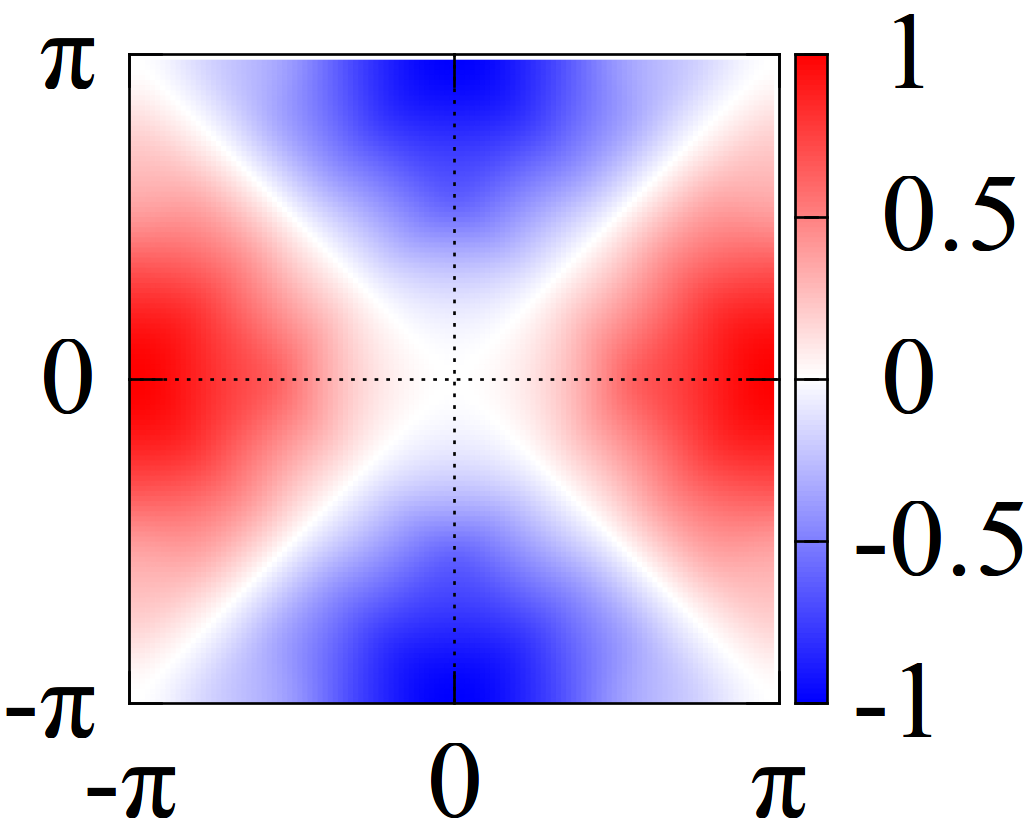}
    \subcaption{n=0.75}
    \label{fig:phi_n=0.75}
        \end{center}
  \end{minipage}
  \begin{minipage}[b]{0.24\linewidth}
    \begin{center}
    \includegraphics[keepaspectratio, scale=0.08]{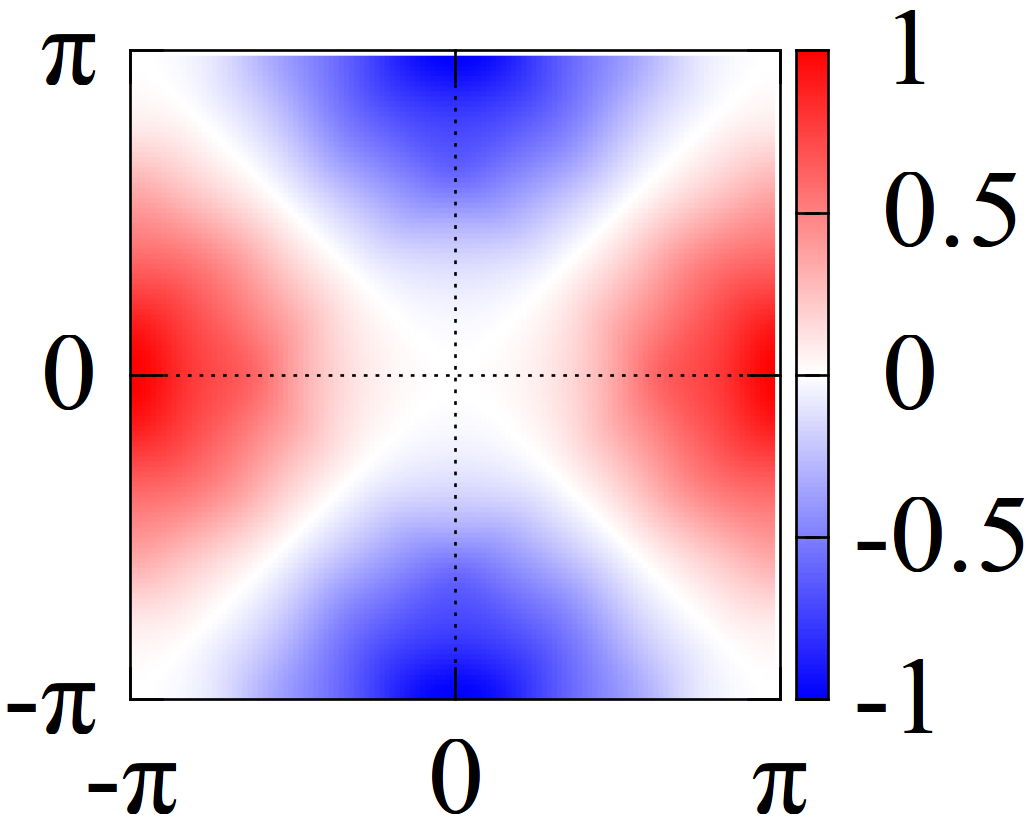}
    \subcaption{n=0.85}
    \label{fig:phi_n=0.85}
        \end{center}
  \end{minipage}
  \begin{minipage}[b]{0.24\linewidth}
    \begin{center}
    \includegraphics[keepaspectratio, scale=0.08]{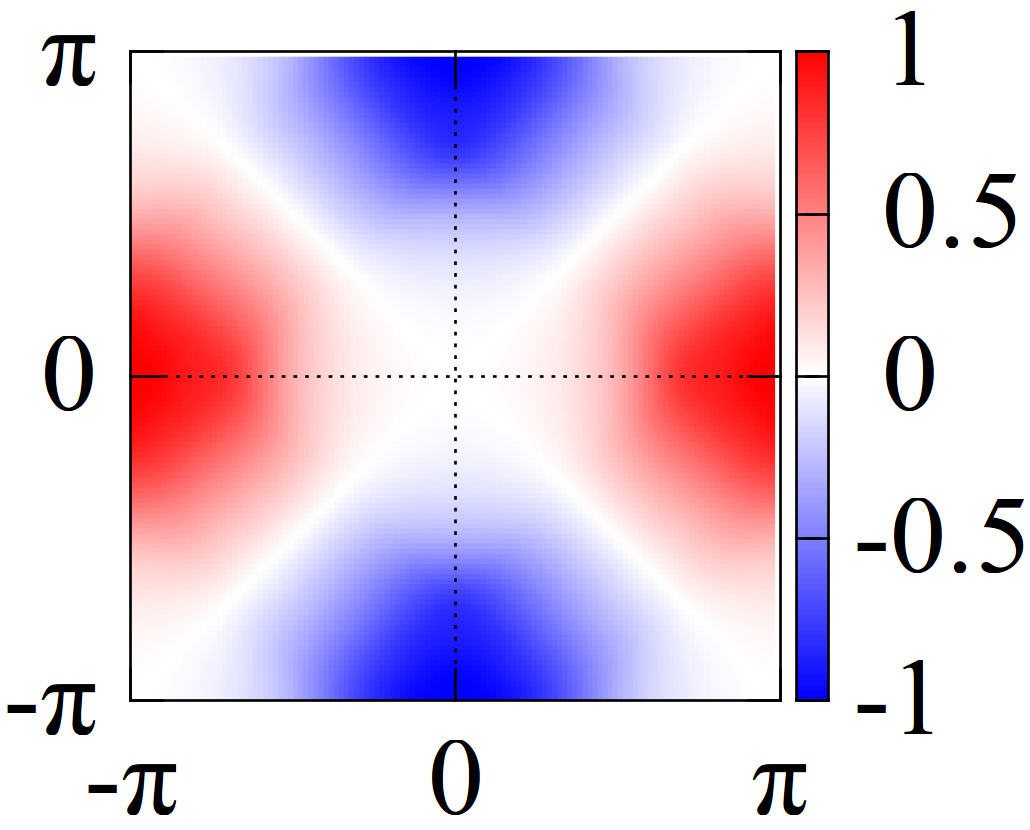}
    \subcaption{n=0.95}
    \label{fig:phi_n=0.95}
        \end{center}
  \end{minipage} \\
  \begin{minipage}[b]{0.24\linewidth}
    \begin{center}
    \includegraphics[keepaspectratio, scale=0.08]{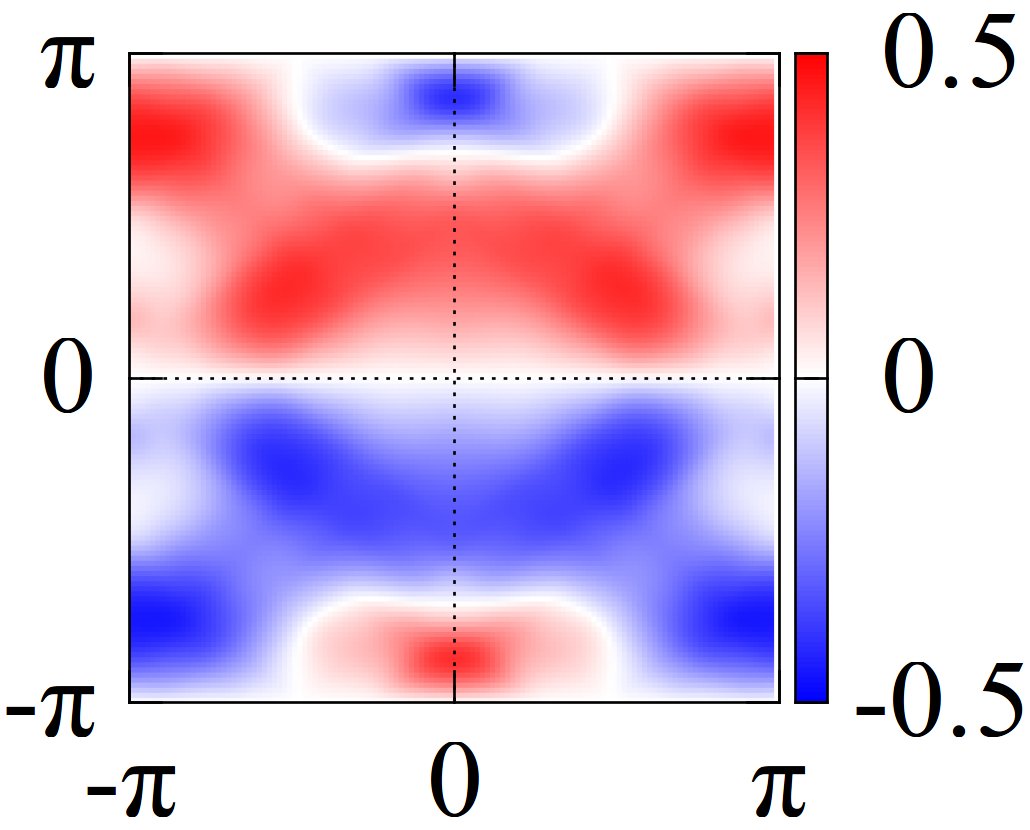}
    \subcaption{n=0.65}
    \label{fig:dx_n=0.65}
        \end{center}
  \end{minipage}
  \begin{minipage}[b]{0.24\linewidth}
    \begin{center}
    \includegraphics[keepaspectratio, scale=0.08]{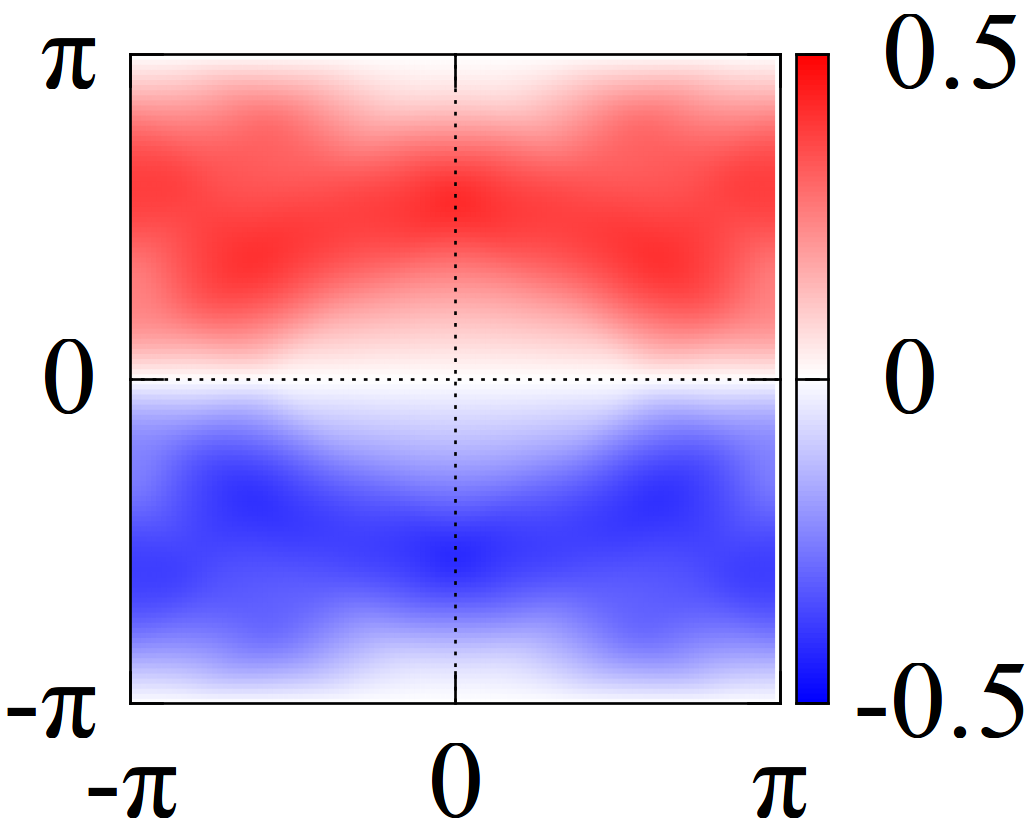}
    \subcaption{n=0.75}
    \label{fig:dx_n=0.75}
        \end{center}
  \end{minipage}
  \begin{minipage}[b]{0.24\linewidth}
    \begin{center}
    \includegraphics[keepaspectratio, scale=0.08]{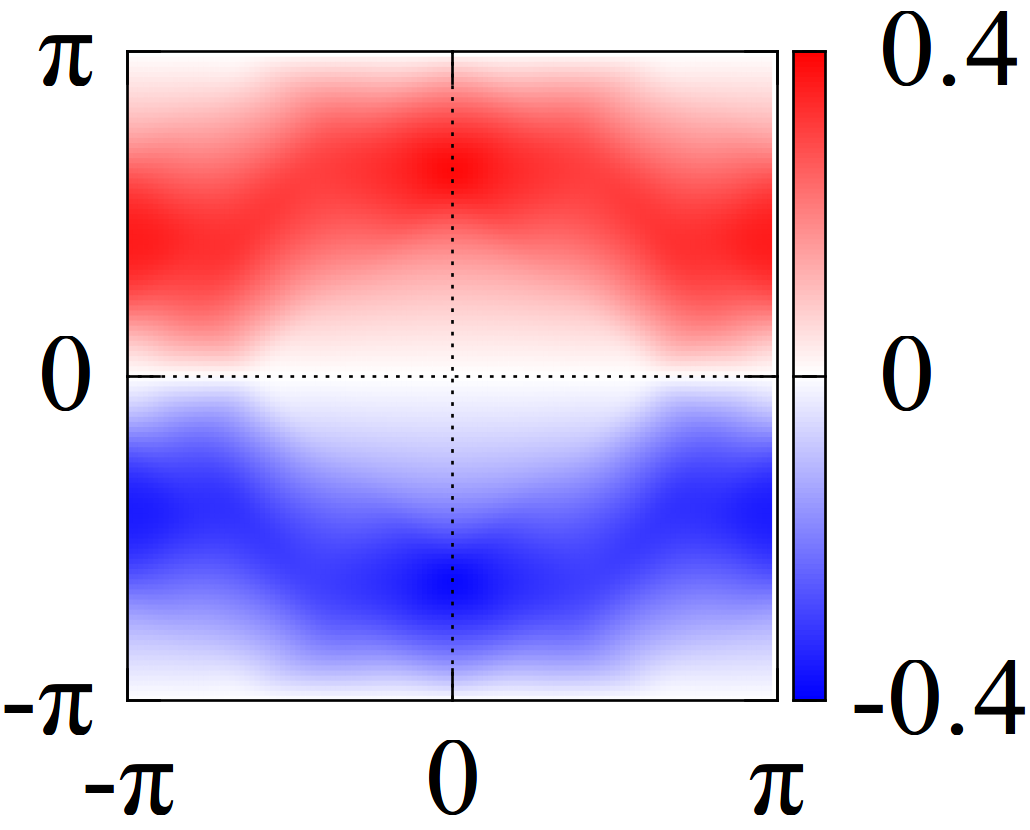}
    \subcaption{n=0.85}
    \label{fig:dx_n=0.85}
        \end{center}
  \end{minipage}
  \begin{minipage}[b]{0.24\linewidth}
    \begin{center}
    \includegraphics[keepaspectratio, scale=0.08]{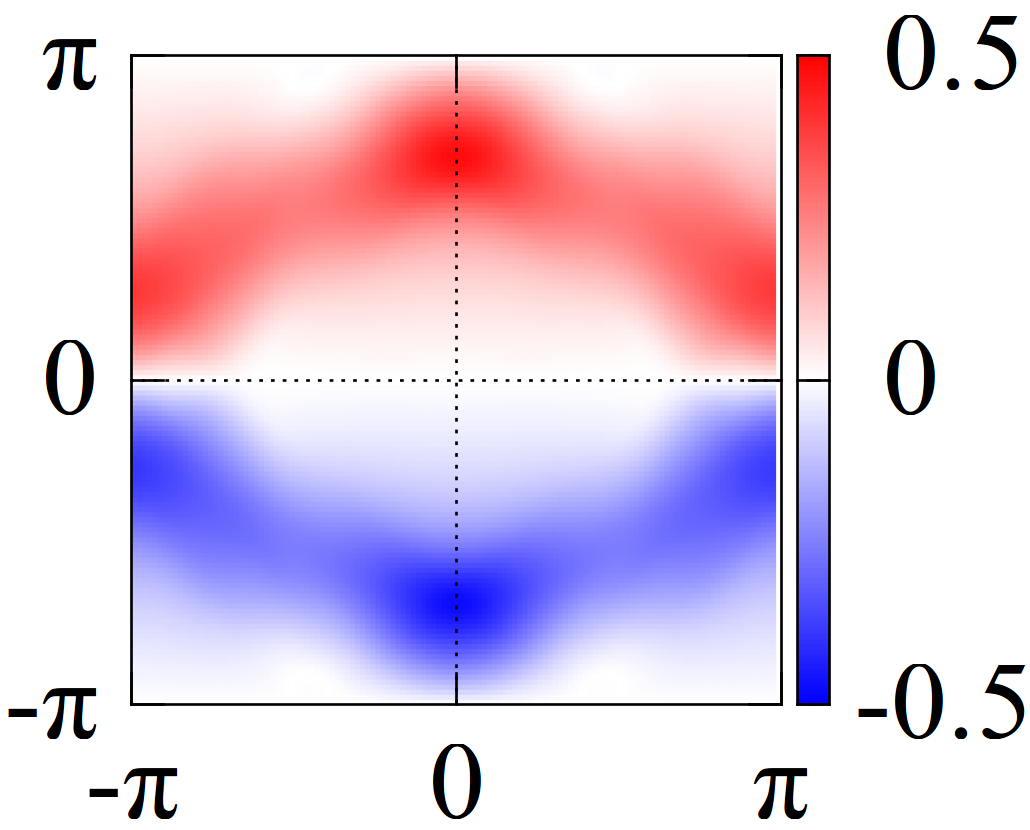}
    \subcaption{n=0.95}
    \label{fig:dx_n=0.95}
        \end{center}
  \end{minipage}
  \begin{minipage}[b]{0.24\linewidth}
    \begin{center}
    \includegraphics[keepaspectratio, scale=0.08]{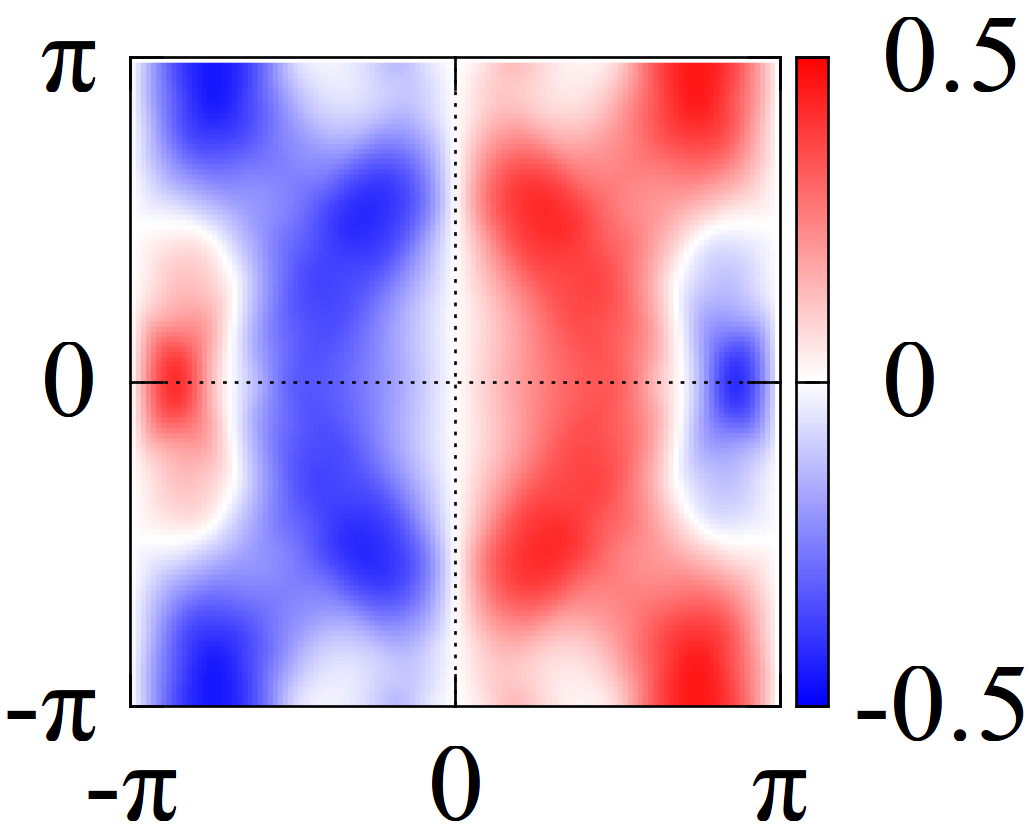}
    \subcaption{n=0.65}
    \label{fig:dy_n=0.65}
        \end{center}
  \end{minipage}
  \begin{minipage}[b]{0.24\linewidth}
    \begin{center}
    \includegraphics[keepaspectratio, scale=0.08]{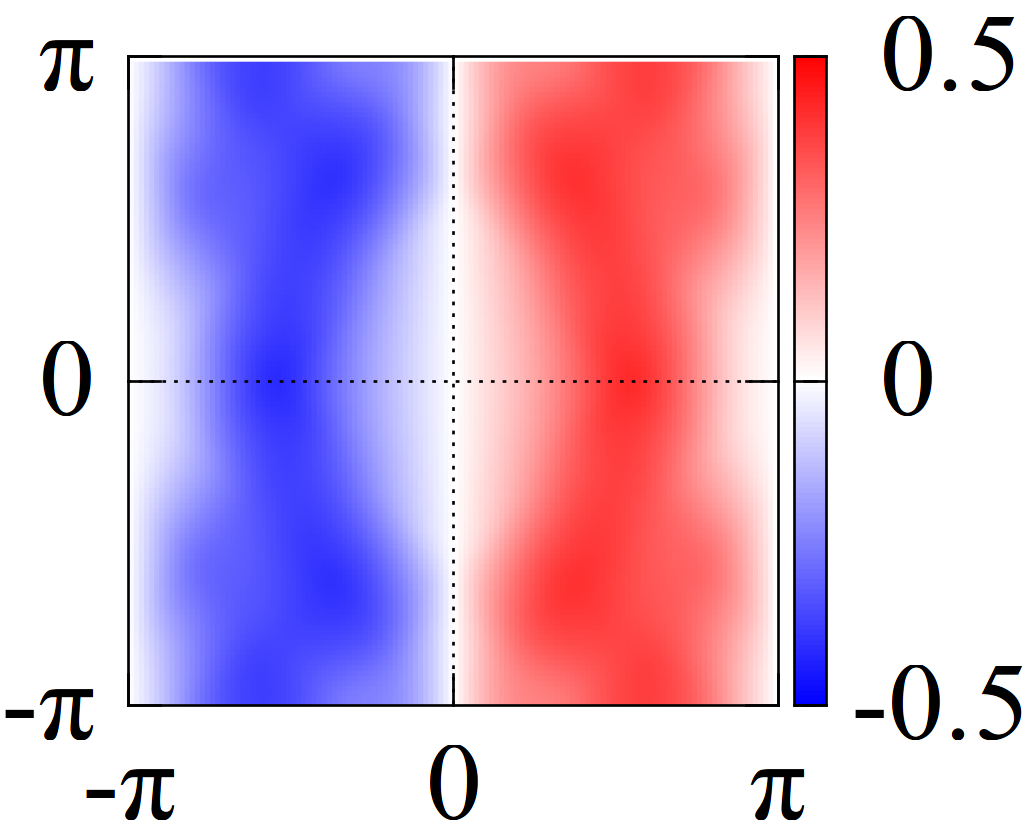}
    \subcaption{n=0.75}
    \label{fig:dy_n=0.75}
        \end{center}
  \end{minipage}
  \begin{minipage}[b]{0.24\linewidth}
    \begin{center}
    \includegraphics[keepaspectratio, scale=0.08]{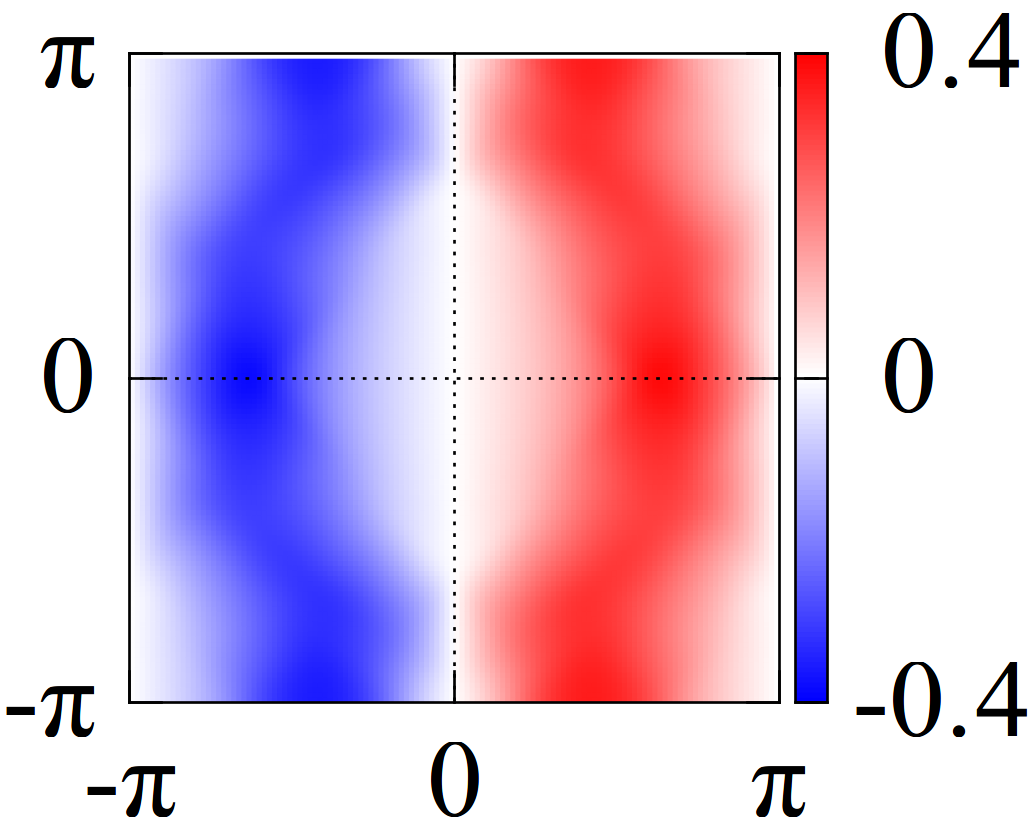}
    \subcaption{n=0.85}
    \label{fig:dy_n=0.85}
        \end{center}
  \end{minipage}
  \begin{minipage}[b]{0.24\linewidth}
    \begin{center}
    \includegraphics[keepaspectratio, scale=0.08]{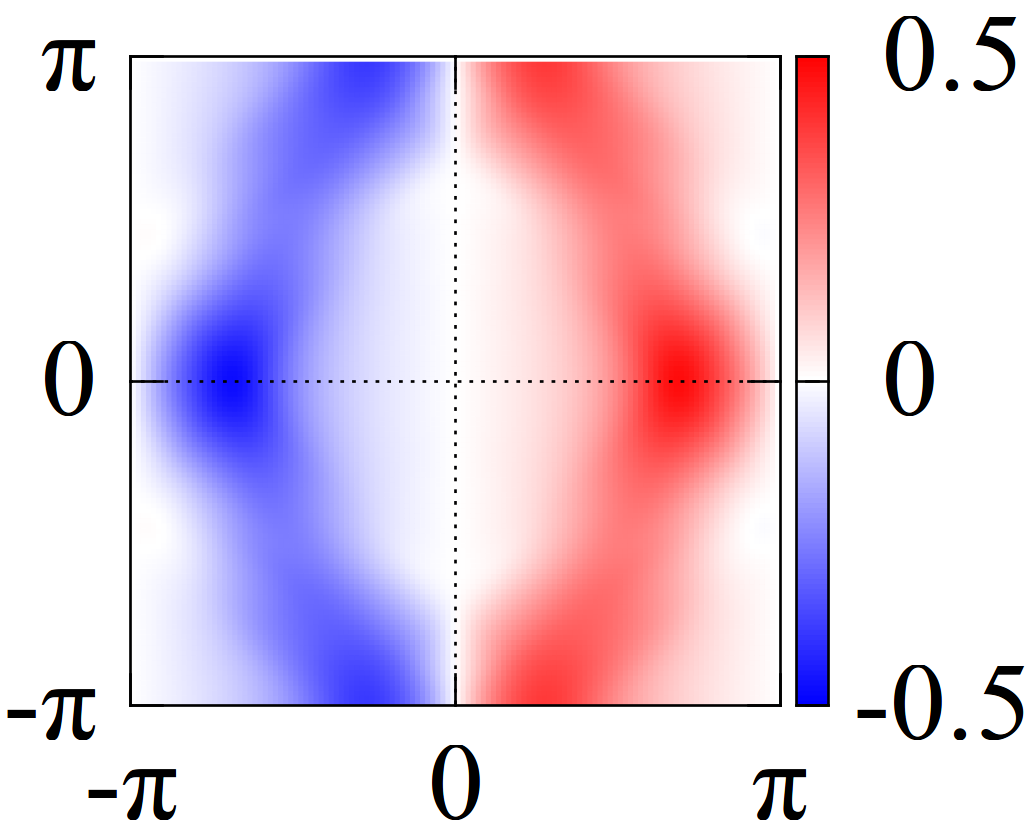}
    \subcaption{n=0.95}
    \label{fig:dy_n=0.95}
        \end{center}
  \end{minipage}
  \caption{(Color online) (\subref{fig:phi_n=0.65})-(\subref{fig:phi_n=0.95}) Spin-singlet component of gap functions, $\psi(\bm{k})$.
    (\subref{fig:dx_n=0.65})-(\subref{fig:dx_n=0.95}) $x$ component and
    (\subref{fig:dy_n=0.65})-(\subref{fig:dy_n=0.95}) $y$ component of the spin-triplet gap functions, namely, the d-vector $\bm{d}(\bm{k})$.
    The parameters are the same as Fig.~\ref{fig:FS_sus}.
    }
  \label{fig:sc_gap}
\end{figure}

The superconducting order parameter of $B_1$ representation contains spin-singlet $d_{x^2-y^2}$-wave pairing as well as spin-triplet pairing with either $p$-wave or $f$-wave symmetry. Because of the Rashba ASOC, superconducting order parameters with distinct space inversion parity coexist. The gap functions are decomposed into the spin-singlet component $\psi(\bm{k})$ and spin-triplet component $\bm{d}(\bm{k})$ in a standard manner, 
\begin{align}
\hat{\Delta}(\bm{k})=\left(\psi(\bm{k})+\bm{d}(\bm{k})\cdot\bm{\sigma}\right)i\sigma_y. 
\end{align}
Fig.~\ref{fig:sc_gap} shows gap functions of the most stable $B_1$ state for various fillings. 
In the whole parameter range, a strongly parity-mixed superconducting state is stabilized. Although the $d_{x^2-y^2}$-wave pairing is always dominant, the subdominant spin-triplet pairing component changes the momentum dependence as a function of the filling. The $d_{x^2-y^2}+f$-wave state is stabilized for $n=0.65$, whereas the $d_{x^2-y^2}+p$-wave state is stable for other fillings. As we have shown in Sec.~\ref{sec:spin_suscep}, the longitudinal spin susceptibilities show qualitatively different behaviors between $n=0.65$ and other fillings. The correspondence between Figs.~\ref{fig:FS_sus} and \ref{fig:sc_gap} implies that the CAFM fluctuation favors the $d_{x^2-y^2}+f$-wave pairing whereas the IAFM fluctuation favors the $d_{x^2-y^2}+p$-wave pairing. 
This is consistent with the previous RPA analysis where the CAFM fluctuation arising from the strong nesting ($t'=0$ and $n \sim 1$) stabilizes a $d_{x^2-y^2}+f$-wave state~\cite{Yokoyama2007}.

\begin{figure}[tbp]
  \begin{minipage}[b]{0.24\linewidth}
    \centering
    \includegraphics[keepaspectratio, scale=0.08]{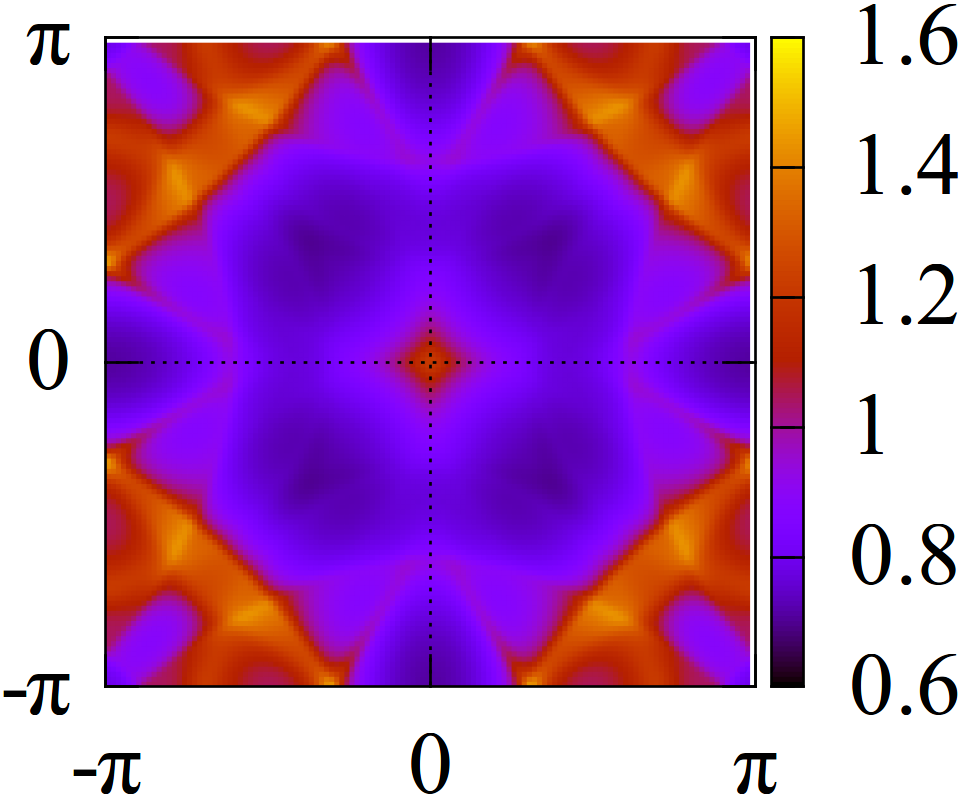}
    \subcaption{$\chi^{zz}(\bm{q},0)$}
    \label{fig:lki_n=0.85_U=2.4}
  \end{minipage}
  \begin{minipage}[b]{0.24\linewidth}
    \centering
    \includegraphics[keepaspectratio, scale=0.08]{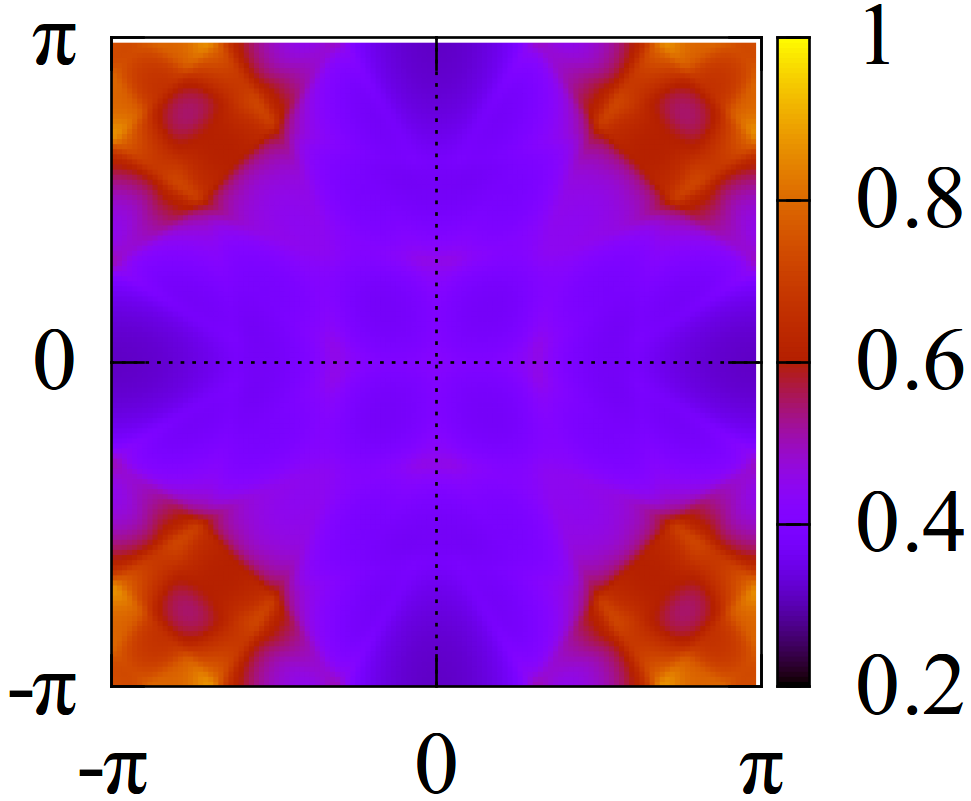}
    \subcaption{$\chi^{-+}(\bm{q},0)$}
    \label{fig:tki_n=0.85_U=2.4}
  \end{minipage}
  \begin{minipage}[b]{0.24\linewidth}
    \centering
    \includegraphics[keepaspectratio, scale=0.08]{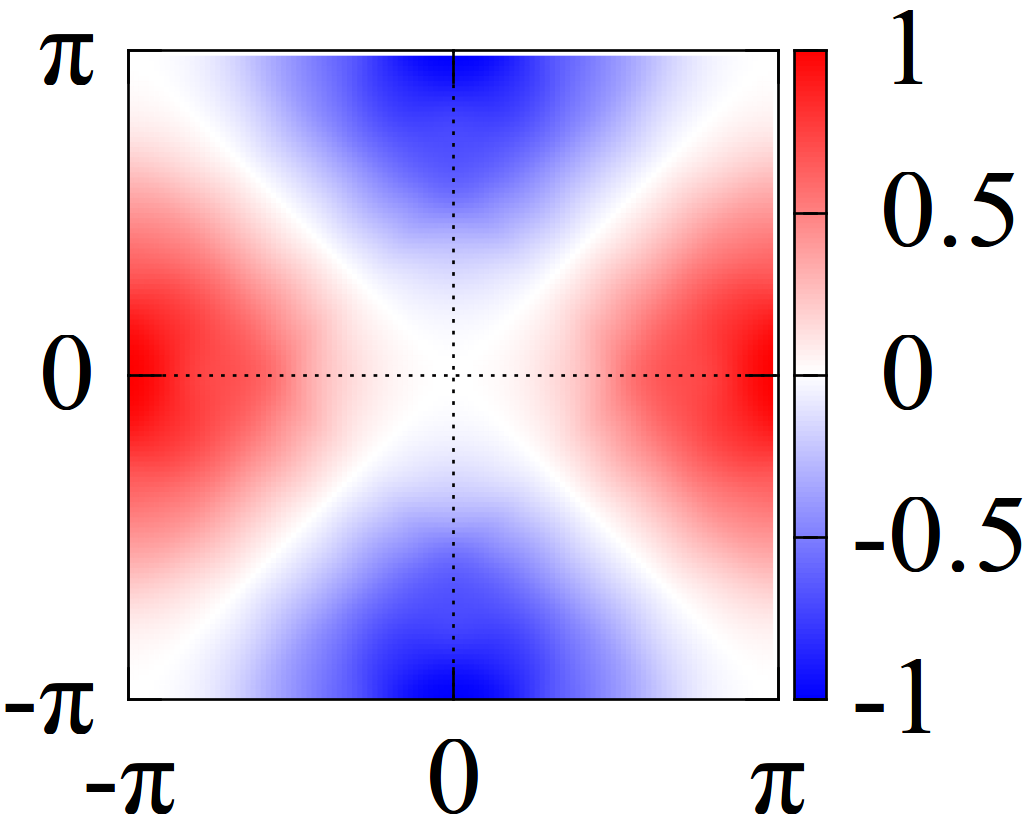}
    \subcaption{$\psi({\bm{k})}$}
    \label{fig:phi_n=0.85_U=2.4}
  \end{minipage}
  \begin{minipage}[b]{0.24\linewidth}
    \centering
    \includegraphics[keepaspectratio, scale=0.08]{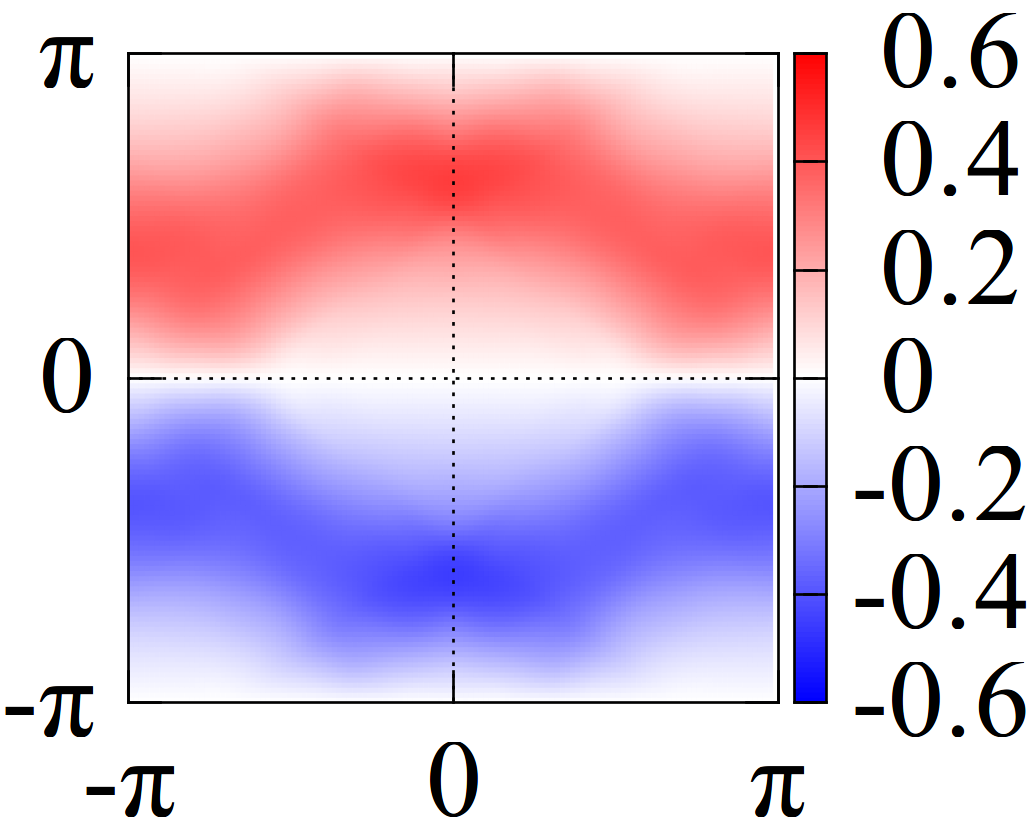}
    \subcaption{$d_x(\bm{k})$}
    \label{fig:dx_n=0.85_U=2.4}
  \end{minipage} \\
  \caption{(Color online) Results for a moderate Coulomb interaction $U=2.4$ and filling $n=0.85$. (\subref{fig:lki_n=0.85_U=2.4}) Longitudinal and  (\subref{fig:tki_n=0.85_U=2.4}) transverse spin susceptibility.
  (\subref{fig:phi_n=0.85_U=2.4}) Spin-singlet component of gap functions, $\psi({\bm{k})}$. (\subref{fig:dx_n=0.85_U=2.4}) $x$ component of the spin-triplet gap functions, $d_x(\bm{k})$.
  }
  \label{fig:U=2.4}
\end{figure}

Here we compare our results with a previous work which investigated similar parameter range within the RPA~\cite{greco2018}. 
The authors of Ref.~\cite{greco2018} claimed that various superconducting states with  different symmetry are stabilized depending on the filling. In particular, the spin-triplet $f$-wave pairing state near the type-II van Hove singularity has been illustrated, and its origin was attributed to the FM spin fluctuation. 
In contrast to their results, our calculation based on the FLEX approximation with the linearized \'{E}liashberg equation shows that the gap functions of superconductivity are essentially independent of the filling and the $d_{x^2-y^2}$-wave paring is dominant. Furthermore, appearance of the subdominant $f$-wave pairing seems to be correlated to the CAFM spin fluctuation rather than the FM spin fluctuation. Indeed, when the FM spin fluctuation clearly appears for a moderate $U=2.4$, the gap functions are almost unchanged, and not the $f$-wave but the $p$-wave pairing is subdominant [see Fig.~\ref{fig:U=2.4}]. Although the instability to the $d_{xy}$-wave superconductivity ($B_{2}$ representation) was also shown in Ref.~\cite{greco2018}, we do not see such tendency [see Fig.~\ref{fig:eigen}(\subref{fig:eigen_n=0.85}) for instance].
Although the origin of different conclusions is unclear and more elaborated studies are desired, it may be partly because the linearized \'{E}liashberg equation is not fully solved in Ref.~\cite{greco2018}.

\begin{figure}[tbp]
  \begin{center}
    \includegraphics[keepaspectratio, scale=0.18]{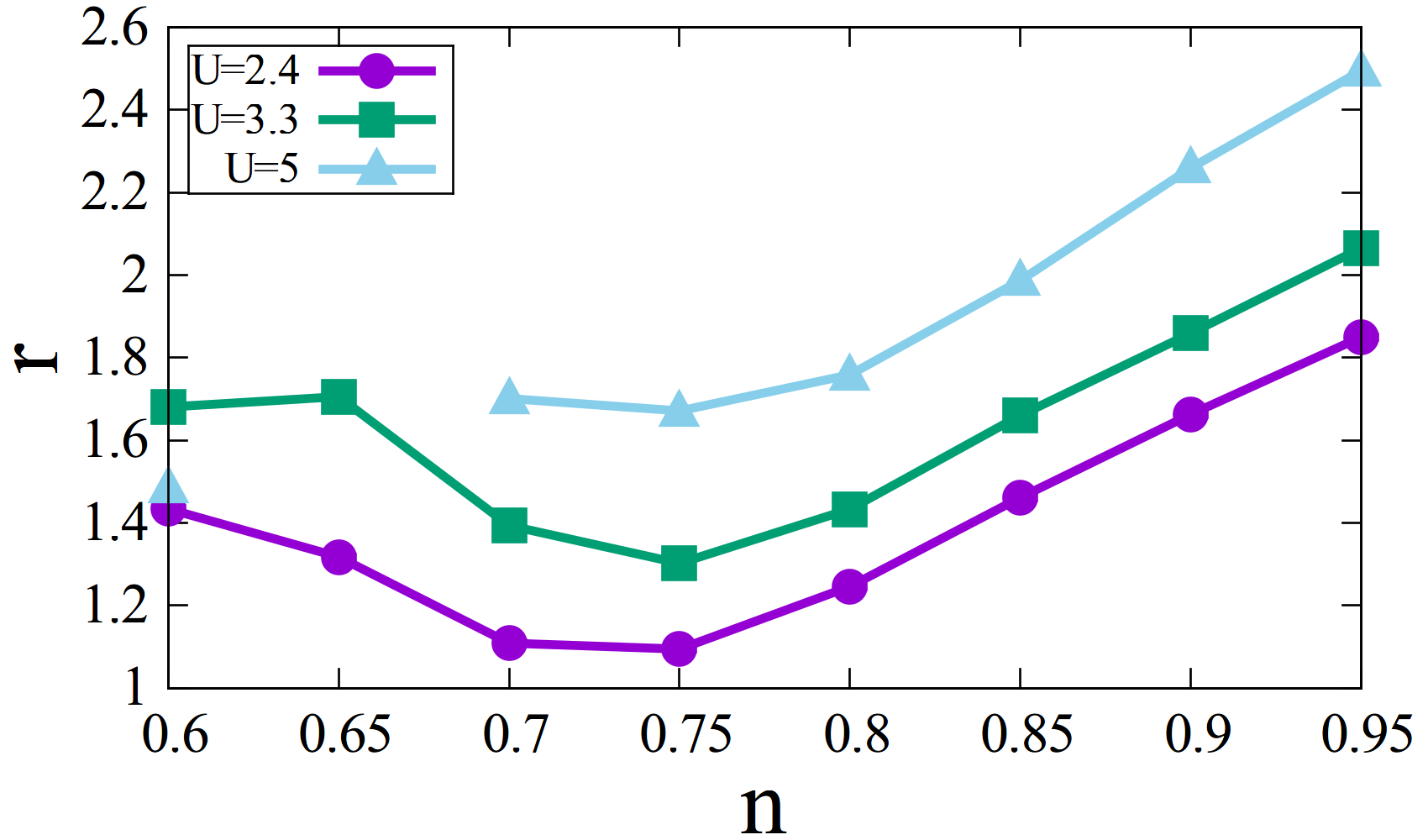}
  \end{center}
  \caption{(Color online) Ratio of spin-singlet paring and spin-triplet pairing, defined in Eq.~\eqref{eq:ratio}.  
  We show the filling dependence for $U=2.4$, $3.3$ and $5$.}
  \label{fig:ratio}
\end{figure}

Because of lack of inversion symmetry, the spin-singlet component and the spin-triplet component coexist in our solution. Although the $d_{x^2-y^2}$-wave superconducting state is extremely stable in the single-band Hubbard model ($\alpha=0$), with a moderate Rashba ASOC $\alpha=0.5$, the magnitude of the spin-triplet component is comparable with the
spin-singlet one. Fig.~\ref{fig:ratio} illustrates the ratio of the magnitudes which is evaluated by 
\begin{equation}
  r=\frac{\sum_{\bm{k}}|\psi(\bm{k})|^2}{\sum_{\bm{k}}|\bm{d}(\bm{k})|^2}.
  \label{eq:ratio}
\end{equation}
We see $r>1$ in the whole parameter range, indicating the dominant $d_{x^2-y^2}$-wave pairing. However, the value of $r$ is close to unity, and therefore, strongly parity-mixed superconducting states are concluded. The parity mixing is particularly enhanced around $n=0.75$, in which the FSs lie between the type-I and type-II van Hove singularity.

\section{conclusions and discussions}
In summary, we have conducted a thorough study on superconductivity in the two-dimensional Rashba-Hubbard model, a minimal model for strongly-correlated noncentrosymmetric electron systems. 
With use of the FLEX approximation combined with linearlized \'{E}liashberg equation, we have clarified interplay of Rashba spin-orbit coupling and critical magnetic fluctuations in a wide range of filling from type-II van Hove singularity to half-filling. 
Our results reveal robust FSs against the critical magnetic fluctuation, enhancement of IAFM fluctuation, and stabilization of strongly parity-mixed superconducting state in a wide parameter range. 
The obtained gap functions show the $d_{x^2-y^2}+p$-wave superconductivity for the filling $n=0.75$, $0.85$, and $0.95$. 
On the other hand, for $n=0.65$, 
we have observed impacts of the type-II van Hove singularity near the Fermi level. The FSs undergo Lifshitz transition due to the electron correlation, the CAFM fluctuation is strongly enhanced, and the $d_{x^2-y^2}+f$-wave superconducting state is stabilized. Strong parity mixing in the gap functions has been observed in the whole parameter range. In particular, magnitude of spin-triplet gap function is comparable to that of spin-singlet one when the Fermi level lies between the type-I and type-II van Hove singularities. 

This work resolved unsettled issues on the strongly correlated Rashba systems~\cite{greco2018,greco2019ferromagnetic,fujimoto2015deformation,maruyama2015}, and elucidated a microscopic mechanism to stabilize a strongly parity-mixed superconducting phase. This finding opens a way to realize intriguing phenomena arising from parity mixing in superconducting order parameters. For instance, as proposed by recent theoretical studies, noncentrosymmetric superconductors with strong parity mixing may be a platform of fractional flux quanta~\cite{iniotakis2008}, nonreciprocal electric current~\cite{wakatsuki2018}, and topological superconductivity~\cite{daido2016,daido2017,takasan2017,lu2018,yoshida2015,yoshida2017}. 

Another class of superconducting phases with strong parity mixing may be stabilized by a critical fluctuation of structural transitions breaking the space inversion symmetry, that is named odd-parity electric multipole fluctuations~\cite{Kozii2015,sumita2020}. In this case, phonons coupled to dynamical spin-orbit coupling mediate pairing interaction in both spin-singlet and spin-triplet channels. 
In contrast, in our proposal anisotropic magnetic fluctuations naturally lead to a strongly parity-mixed superconducting state in quasi-two-dimensional electron systems with strong electron correlations. The candidates may be naturally-formed or artificially-engineered heterostructures of cuprate superconductors~\cite{bollinger2011,Gotlieb1271} or heavy fermion superconductors~\cite{mizukami2011extremely,Goh2012,Shimozawa2014,Shimozawa_2016,naritsuka2017,naritsuka2018}.

\begin{acknowledgments}
 The authors are grateful to J.~Ishizuka, A.~Daido, S.~Sumita, and H.~Watanabe for fruitful discussions.
 This work was supported by JSPS KAKENHI (Grants No. JP15H05884, No. JP18H04225, No.
JP18H05227, No. JP18H01178, and No. 20H05159).
The numerical calculations were carried out on Cray xc40 at YITP in Kyoto University.
\end{acknowledgments}


%

\end{document}